\documentclass[journal]{IEEEtran}

\usepackage[bookmarks=false]{hyperref}
\hypersetup{
    colorlinks=true,
    linkcolor=black,
    filecolor=black,
    urlcolor=black,
}

\usepackage{algpseudocode}
\usepackage[utf8]{inputenc}
\usepackage[T1]{fontenc}
\usepackage{algorithm} 
\usepackage{nomencl}
\usepackage[normalem]{ulem}
\usepackage{etoolbox}
\usepackage[dvipsnames]{xcolor}
\usepackage{wrapfig}
\usepackage{graphicx}
\graphicspath{ {./images/} }

\newcommand{\maybe}[1]{}

\usepackage{tikz}
\usetikzlibrary{matrix}

\makeatletter
\newdimen\multi@col@width
\newdimen\multi@col@margin
\newcount\multi@col@count
\tikzset{
  multicol/.code={%
    \global\multi@col@count=#1\relax
    \global\let\orig@pgfmatrixendcode=\pgfmatrixendcode
    \global\let\orig@pgfmatrixemptycode=\pgfmatrixemptycode
    \def\pgfmatrixendcode##1{\orig@pgfmatrixendcode%
      ##1%
      \pgfutil@tempdima=\pgf@picmaxx
      \global\multi@col@margin=\pgf@picminx
      \advance\pgfutil@tempdima by -\pgf@picminx
      \divide\pgfutil@tempdima by #1\relax
      \global\multi@col@width=\pgfutil@tempdima
      \pgf@picmaxx=.5\multi@col@width
      \pgf@picminx=-.5\multi@col@width
      \global\pgf@picmaxx=\pgf@picmaxx
      \global\pgf@picminx=\pgf@picminx
      \gdef\multi@adjust@position{%
        \setbox\pgf@matrix@cell=\hbox\bgroup
        \hfil\hskip-\multi@col@margin
        \hfil\hskip-.5\multi@col@width
        \box\pgf@matrix@cell
        \egroup
      }%
      \gdef\multi@temp{\aftergroup\multi@adjust@position}%
      \aftergroup\multi@temp
    }
    \gdef\pgfmatrixemptycode{%
      \orig@pgfmatrixemptycode
      \global\advance\multi@col@count by -1\relax
      \global\pgf@picmaxx=.5\multi@col@width
      \global\pgf@picminx=-.5\multi@col@width
      \ifnum\multi@col@count=1\relax
       \global\let\pgfmatrixemptycode=\orig@pgfmatrixemptycode
      \fi
    }
  }
}
\makeatother
\usepackage{blindtext}
\usepackage{multicol}
\ifCLASSINFOpdf
  \graphicspath{{../pdf/}{../jpeg/}}
  \DeclareGraphicsExtensions{.pdf,.jpeg,.png}
\else

\fi

\usepackage{cite}
\usepackage{amsmath,amssymb,amsfonts}
\usepackage{algorithm}
\usepackage{acro}
\usepackage{algpseudocode}
\usepackage{graphicx}
\usepackage{textcomp}
\usepackage{xcolor}
\usepackage{tikz}
\usepackage{adjustbox}
\usepackage[version=4]{mhchem}
\usepackage{booktabs}
\usepackage{multirow}
\usepackage{siunitx}
\usepackage{eurosym}
\usepackage{mhchem}
\usepackage{amssymb}
\usepackage{tikz}
\usepackage{pgfplots}
\usepackage{multirow}
\usepackage{graphicx}
\usepackage{hyperref}
\usepackage{eurosym}
\usepackage{subcaption}
\def\BibTeX{{\rm B\kern-.05em{\sc i\kern-.025em b}\kern-.08em
    T\kern-.1667em\lower.7ex\hbox{E}\kern-.125emX}}
\usepackage{url}

\newtheorem{definition}{Definition} 

\DeclareAcronym{is}{
  short=IS,
  long=In-sample,
}
\DeclareAcronym{oos}{
  short=OOS,
  long=Out-of-sample,
}
\DeclareAcronym{system operator}{
  short=system operator,
  long=Transmission System Operator,
}
\DeclareAcronym{system operators}{
  short=system operators,
  long=Transmission System Operators,
}
\DeclareAcronym{jcc}{
  short=JCC,
  long=Joint Chance-Constraint,
}
\DeclareAcronym{jccp}{
  short=JCCP,
  long=Joint Chance-Constrained Program,
}

\DeclareAcronym{fcrd}{
  short=FCR-D,
  long= Frequency Containment Reserve - Disturbance,
}

\DeclareAcronym{fcrn}{
  short=FCR-N,
  long=Frequency Containment Reserve - Normal operation
}
\DeclareAcronym{mfrr}{
    short=mFRR,
    long=Manual Frequency Restoration Reserve
}

\DeclareAcronym{ler}{
  short=LER,
  long=Limited Energy Reservoir
}

\DeclareAcronym{ev}{
  short=EV,
  long=Electric Vehicle
}
\DeclareAcronym{cvar}{
  short=CVaR,
  long=Conditional Value at Risk
}
\DeclareAcronym{var}{
  short=VaR,
  long=Value at Risk
}
\DeclareAcronym{soc}{
  short=SoC,
  long=State of Charge
}
\DeclareAcronym{lp}{
  short=LP,
  long=Linear Program
}
\DeclareAcronym{milp}{
  short=MILP,
  long=Mixed-Integer Linear Program
}

\begin{document}
%

\title{Aggregator of Electric Vehicles Bidding in Nordic FCR-D Markets: A Chance-Constrained Program}


\author{Gustav A. Lunde, Emil V. Damm, Peter A.V. Gade, and Jalal Kazempour, \textit{Senior~Member,~IEEE}

\vspace{-5mm}

\thanks{
This work  was supported in part by Innovation Fund
Denmark under grant number 0153-00205B.  G. A. Lunde, E. V. Damm, P. A. V Gade, and J. Kazempour  are with the Department of Wind and Energy Systems, Technical University of Denmark, Kgs. Lyngby
2800, Denmark (e-mails: Lunde.gustav@gmail.com, ev.damm@yahoo.com, \{pega, jalal\}@dtu.dk). P. A. V Gade is also with IBM Client Innovation Center, Copenhagen, Denmark. The first two co-authors contributed  equally.  
}
}

    




\vspace{-5mm}
\maketitle


\IEEEaftertitletext{\vspace{-0.8\baselineskip}}
\maketitle
\thispagestyle{plain}
\pagestyle{plain}
\begin{abstract}
The Danish system operator, Energinet, has recently introduced an innovative grid code called the $\rm{P90 \  requirement}$, which allows stochastic flexible resources to bid their flexibility in Nordic ancillary service markets, contingent upon a minimum 90\% probability of successfully realizing the reserve capacity bid. For limited-energy resources, Energinet imposes additional requirements for participation in these markets.
Given these requirements, this paper presents a chance-constrained optimization model designed for aggregators of electric vehicles, aiming to optimally place reserve capacity bids in the Nordic Frequency Containment Reserve for Disturbances (FCR-D) market while accounting for uncertainty in future consumption baselines. We analyze both FCR-D up and down markets, reformulating and solving the proposed joint chance-constrained model using two sample-based methods.
Using real data from 1400 charging stations in Denmark from March 2022 to March 2023, we demonstrate the out-of-sample profit potential. Our findings indicate that vehicle owners could save between 6\% and 10\% on their annual electricity bills by providing FCR-D services. Additionally, we observed a synergy effect, where having more vehicles in a single portfolio enables larger bids per vehicle compared to a collective bid from multiple portfolios with the same total number of vehicles.

\end{abstract}
\vspace{1mm}


\vspace{1mm}
\section{Introduction}\label{sec:Introduction}


\subsection{Background and motivation}


A new grid code introduced by the Danish system operator, Energinet, known as the ``$\rm{P90}$ requirement'' \cite{energinet2023prequalification}, addresses how stochastic flexible resources can contribute to frequency-supporting ancillary services by incorporating a degree of uncertainty in their reserve capacity bids. Examples of these resources include wind turbines, heat pumps, supermarket freezers, and grid-connected electric vehicles (EVs). Under the $\rm{P90}$ requirement, these flexible resources can participate in the Nordic ancillary service markets, as long as their capacity bids have at least a $90$\% probability of being successfully realized.

To the best of our knowledge, this is the first time that a system operator implements a grid code with a probabilistic measure, explicitly defining a minimum reliability threshold for the provision of reserve services. The $\rm{P90}$ requirement has the potential to enhance the volume of supply and therefore the liquidity of the Nordic ancillary service markets by reducing technical and market barriers for stochastic and distributed flexible resources. Similar efforts, albeit without specifying an explicit probability threshold and primarily focusing on batteries, have been made by the Federal Energy Regulatory Commission (FERC) through Orders $755$ \cite{755} and $784$ \cite{784} in U.S. markets. These orders mandate that independent system operators and regional transmission organizations implement pay-for-performance frequency regulation markets and account for battery constraints in their regulation dispatch models.

Other system operators particularly in Europe have also initiated efforts to attract flexibility from stochastic resources. For example, some are developing alternative (secondary) auctions for reserve capacity, allowing reserve providers to transfer their obligations if they become uncertain about their ability to deliver the service. While this approach enables providers to manage their risks associated with reserve delivery, it largely maintains that risk with the providers themselves. In contrast, Energinet's $\rm{P90}$ requirement acknowledges some of this risk by accepting a certain level of uncertainty in reserve capacity\footnote{This paper will not discuss the impact of the $\rm{P90}$ requirement on Energinet’s reserve dimensioning, specifically the amount of additional reserve capacity that needs to be procured to ensure grid security.}.

\vspace{-3mm}
\subsection{Research question and framework}

The primary research question is how an aggregator of stochastic flexible resources can optimally bid in the reserve market while adhering to the $\rm{P90}$ requirement. We focus on EV aggregators, without loss of generality, as an example of stochastic flexible resources, given their demonstrated capability to provide fast automatic reserve services \cite{hrvoje}. We examine how the $\rm{P90}$ requirement influences their bidding decisions and explore whether aggregators with a larger number of vehicles can leverage this requirement to place larger reserve bids per vehicle.

We consider the flexibility of the EV aggregator to be realized \textit{solely} through adjustments to the total consumption level of grid-connected EVs. Vehicle-to-grid technologies, which enable power injection from the battery to the grid, are excluded from this study. The maximum potential for upwards flexibility (reducing power consumption) is time-variant and depends on the total consumption of the EV fleet. Similarly, the maximum potential for downwards flexibility (increasing power consumption) is also time-variant, as it is influenced by the number of grid-connected EVs at any given time. Both upwards and downwards flexibility potentials are estimated using historical data, similar to other stochastic flexible demands.
Therefore, the EV aggregator is viewed more as a stochastic flexible demand resource rather than a large-scale stationary battery. We assume that the location of EVs does not impact their service provision, and their capability is not constrained by local distribution grid limitations.

Among the various Nordic ancillary services, we focus on a specific service called the Frequency Containment Reserve for Disturbances (FCR-D), which is activated during operational circumstances involving extreme frequency deviations—specifically when the frequency drops below $49.9$ Hz or exceeds $50.1$ Hz. Energinet procures certain amounts of FCR-D services for each hour through a market mechanism, clearing the FCR-D market in a day-ahead time frame, which consists of two distinct segments: one for up services and another for down services. In contrast, such a market does not exist in the Continental Europe synchronous area, where extreme frequency deviations are less likely to occur.

Our focus on the FCR-D market is driven by significant price increases observed in Denmark over the past few years, culminating in a peak price of $16531$ DKK/MW (approximately €$2215$/MW) for FCR-D down services in early $2024$. This surge has triggered considerable investments in technologies capable of participating in the market, and demand-side flexibility presents a particularly viable option, as it requires no additional investments due to the $\rm{P90}$ requirement that allows stochastic demand-side assets to bid.

\subsection{Literature review, contributions, and findings}

First, we would like to emphasize that, to the best of our knowledge, there is currently no literature addressing the $\rm{P90}$ requirement or similar grid codes that incorporate a probabilistic measure. However, several papers tackle related research questions, notably \cite{dirk}, which develops a robust optimization framework that incorporates European legislation for frequency regulation. The central idea of that work is to provide guarantees for delivering regulation power across various frequency deviation scenarios within a defined uncertainty set. In contrast, this work specifically addresses Energinet’s $\rm{P90}$ requirement, which falls within the realm of chance-constrained optimization by specifying a minimum probability of $90$\% for the successful realization of bids. Additionally, other papers in the literature discuss reserve provision by EV aggregators, such as \cite{azi1, azi2, azi3, azi4}, but none consider a grid code that allows for a certain probability of reserve shortfall.

From a methodological perspective, several papers in the literature utilize chance-constrained programming for reserve decision-making. For instance, \cite{a} develops a chance-constrained reserve dimensioning model that enables a system operator to make informed decisions about the quantity of reserve services to procure. Reference \cite{b} defines a probabilistic reserve service for a system operator, while \cite{j,k} present chance-constrained models that assist system operators in optimally dispatching flexible resources across various reserve markets.
From the aggregator's perspective, \cite{d} proposes a distributionally robust chance-constrained model for strategic bidding in a generic reserve market. Similarly, \cite{f} develops an offering strategy model for batteries bidding in frequency regulation markets, incorporating a chance constraint to meet the performance requirements of FERC. A comparable model for batteries is introduced in \cite{h}, where a chance-constrained optimization provides a probabilistic guarantee for reserve availability in real-time scenarios.
However, none of these studies specifically address the requirements of Nordic ancillary service markets. In a recent publication by the authors \cite{e}, a stylized distributionally robust chance-constrained model is developed, but it omits Energinet’s additional requirements regarding limited-energy resources, FCR-D down services, and the synergy effect of aggregating stochastic flexible resources within the same portfolio.

Given this literature review, the contributions of this paper are as follows: We develop an offering strategy model for an EV aggregator with a stochastic consumption baseline, enabling optimal reserve capacity bids in the Nordic FCR-D up and down markets while fulfilling the requirements set by Energinet. This leads to an optimization problem with joint chance constraints. Our contribution primarily focuses on the application aspect—specifically, the integration of a new grid code into the bidding optimization problem—rather than the development of a novel solution methodology. To solve the resulting joint chance-constrained program, we employ two well-established sample-based methods. The first method utilizes an iterative ALSO-X algorithm \cite{Ahmed201751}, which solves a linear program at each iteration. The second method applies a Conditional Value-at-Risk (CVaR) \cite{rockafellar2000optimization} approximation of the joint chance constraints, also resulting in a linear program.

Our findings are as follows: Utilizing real data from $1400$ EV charging boxes in Denmark from March $2022$ to March $2023$, we compare the out-of-sample performance of both methods and find satisfactory results. However, the CVaR method appears to be more conservative---perhaps unnecessarily so from the aggregator's perspective---compared to the ALSO-X method. Our numerical analysis indicates that when the aggregator shares all benefits among EV owners, a typical EV owner could save between $6$\% and $10$\% on average in their annual electricity bill by providing FCR-D services. Additionally, we observe a synergy effect among EVs within the portfolio, demonstrating that having more EVs in a single portfolio enables the aggregator to place a larger reserve capacity bid per EV compared to a collective bid from multiple portfolios with the same total number of EVs.

The remainder of the paper is organized as follows: Section II provides an overview of the Nordic FCR-D up and down services and the corresponding markets. Section III outlines Energinet's requirements and presents the proposed joint chance-constrained bidding problem. Section IV explains the methodology for estimating the available flexibility and drawing samples. Building on these samples, Section V introduces two sample-based methods for solving the proposed joint chance-constrained model. Section VI discusses the numerical results, and finally, Section VII concludes the paper.

\vspace{2mm}
\section{Preliminaries: FCR-D}\label{sec:overview}
The Nordic system operators, including Energinet, procure FCR-D services in advance for activation during extreme frequency deviations from the nominal rate of $50$ Hz. FCR-D up services are activated when the frequency falls within the range of $49.9$ Hz to $49.5$ Hz. The activation percentage depends linearly on the frequency drop: $0$\% activation at $49.9$ Hz and $100$\% activation at $49.5$ Hz. Similarly, FCR-D down services are procured for events when the frequency ranges from $50.1$ Hz to $50.5$ Hz, with the activation level also depending linearly on the frequency\footnote{For more detailed information on the amount of FCR-D up and down services (in MW) procured by each Nordic system operator, and the projected increase in the need for these services in Denmark by $2030$–$2040$, we refer the interested reader to \cite{outlook} and \cite{marco}.}. Under normal operational conditions, when the frequency lies between $49.9$ Hz and $50.1$ Hz, the Nordic system operators activate another service known as FCR for normal operation (FCR-N), which is outside the scope of this paper. It should be noted that FCR-N services are activated much more frequently than FCR-D services.

\begin{figure}[t!]
    \centering
    \includegraphics[width=1\columnwidth]{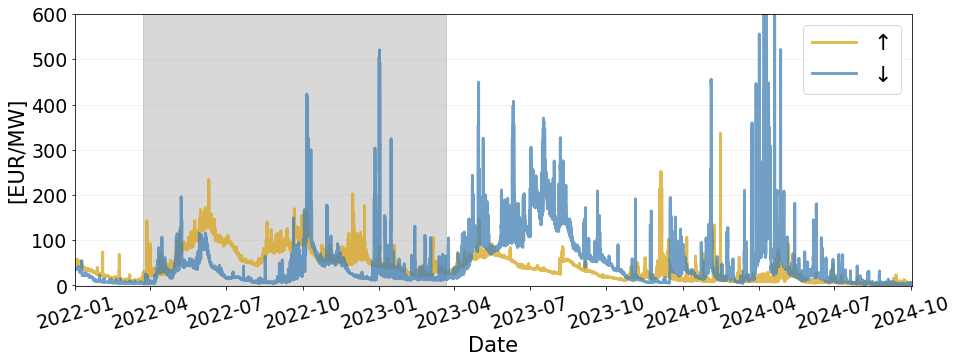}
    \caption{\small{Prices for the FCR-D up ($\uparrow$) and down ($\downarrow$) markets in Denmark from January $1$, $2022$, to October $1$, $2024$ \cite{EnergiDataServiceFCRNDK2}. The maximum price reached €$2215$/MW. In our case study, we focus on the time period highlighted in grey.}}
        \vspace{-1mm}
    \label{fig:prices}
\end{figure}

FCR-D service providers must respond within $2.5$ seconds of activation \cite{energinet2023prequalification}. Since FCR-D events are rare and typically short in duration, the total energy delivery during an activation is minimal. For instance, from March $24$, $2022$, to March $21$, $2023$, the total energy delivered by $1$ MW of FCR-D up and $1$ MW of FCR-D down services in Denmark was $3.25$ MWh and $4.05$ MWh, respectively. These values represent only $0.037$\% and $0.046$\% of the total FCR-D up and down bid capacities, respectively, which is negligible\footnote{The historical frequency data used in this paper comes from \cite{fingrid_frequency_historical_data_2023}, which has been down-sampled to a one-minute resolution based on the maximum and minimum grid frequencies recorded at the millisecond level during that minute.}. The rare activation and minimal energy delivery are two key reasons why Nordic FCR-D services are particularly attractive to assets such as batteries and EV aggregators. Combined with the exclusion of vehicle-to-grid technologies, these factors ensure that providing FCR-D services via EVs does not lead to significant battery degradation or reduce their lifespan.

The historical hourly reservation prices in Denmark for both FCR-D up and down services from January $1$, $2022$, to October $1$, $2024$, are shown in Fig. \ref{fig:prices}, revealing significant volatility and high price levels. Previously, the FCR-D up and down markets operated under a pay-as-bid pricing scheme, but this changed to a uniform scheme in February $2024$. Consequently, Fig. \ref{fig:prices} presents volume-weighted average prices for procured services up to February $2024$, with uniform prices reflected thereafter. Notably, there are prolonged periods of extremely high prices, such as the FCR-D down prices observed throughout much of $2023$. The FCR-D up and down markets consist of early and late auctions, both cleared the day before operation ($\rm{D\!-\!1}$). 
As illustrated in Fig. \ref{fig:timeline_FCR}, we focus on the early auction for simplicity, where reserve capacity bids for FCR-D up and down services, denoted as $c_{h}^{\uparrow}, c_{h}^{\downarrow}$ (in MW), are placed for each hour 
$h$ of the following day ($\rm{D}$). The decisions regarding these hourly bids on day $\rm{D\!-\!1}$ are based on the available forecasts of future stochastic consumption baselines. These forecasts allow the aggregator to generate scenarios $\omega\!\in\!\Omega$ of the available flexibility $F_{m,h,\omega}$ (in MW)  of the EV portfolio for every minute $m$ of hour $h$. We utilize a minute-level resolution based on the available historical data for this study; however, if higher-resolution data (e.g., at the second level) is available, we recommend using it. Our proposed model is flexible enough to accommodate such higher-resolution data.

On the operation day $\rm{D}$, once activated, energy $A_{m,h}$ (in MWh) must be delivered according to the realized frequency deviation and the available flexibility of the EV portfolio. FCR-D service providers are remunerated ex-post at the FCR-D reservation prices $\pi_h$ (in DKK/MW) and are penalized for any missed reserve $p_{h}^{\uparrow}, p_{h}^{\downarrow}$ during activation at the penalty price $\lambda_{h}$ (in DKK/MW).

\begin{figure}[t]
    \begin{adjustbox}{width=\columnwidth}
        \definecolor{RectangleColor}{RGB}{144,238,144} 

\begin{tikzpicture}[scale=0.7,every text node part/.style={align=center}]

  \definecolor{RectangleColor}{RGB}{10,163,10} 

  \draw[rounded corners, fill=RectangleColor, very thick] (0, 1) rectangle (4, 2.5) node[midway]{FCR-D (early) \\reservation bids};
  \node at (2.4,0) {$c_{h}^{\uparrow}, c_{h}^{\downarrow}$ \quad $(F_{m,h,\omega})$};
  \node at (2,-1) {D$-$1};
  \node at (2,-1.6) {(0:30am)};
  
  \draw[rounded corners, fill=RectangleColor, very thick] (8, 1) rectangle (12, 2.5) node[midway]{FCR-D \\ activation};
  \node at (10,0) { \quad ($A_{m,h},F_{m,h}$) };
  \node at (10,-1) {D};
  \node at (10,-1.6) {(12pm -- 12am)};
  
  \draw[rounded corners, fill=RectangleColor, very thick] (16, 1) rectangle (20, 2.5) node[midway]{Ex-post \\ Settlement};
  \node at (18,0) {$p_{h}^{\uparrow}, p_{h}^{\downarrow}$  \quad  $(\pi_{h}, \lambda_{h})$};
  \node at (18,-1) {D+1};
  \node at (18,-1.6) {(12pm)};

  \draw[->, line width=0.65mm, black] (-1, -0.5) -- (21, -0.5);

  \draw[dashed, line width=0.65mm, black] (6, -0.5) -- (6, 0.5);
  \draw[dashed, line width=0.65mm, black] (14, -0.5) -- (14, 0.5);

\end{tikzpicture}

    \end{adjustbox}
    \caption{\small{Timeline of FCR-D up and down markets in Denmark. We focus on the early FCR-D market only. The optimization variables are denoted by lower-case letters, while upper-case and Greek letters are used for parameters.
    }
    }
    \label{fig:timeline_FCR}
\end{figure}

\vspace{1mm}
\section{Energinet's requirements and \\ the resulting bidding problem}\label{sec:math}

In this section, we first outline Energinet's requirements and then develop a bidding decision-making model for the EV aggregator, formulated as a joint chance-constrained optimization problem.

\vspace{4mm}
\subsection{The $\rm{P90}$ requirement}
Energinet defines the $\rm{P90}$ requirement as follows \cite{energinet2023prequalification}:

\begin{definition}[The $\rm{P90}$ requirement]\label{def:P90}
    \textit{``[...] This means, that the participant's prognosis, which must be approved by Energinet, evaluates that the probability is 10\% that the sold capacity is not available. This entails that there is a 90\% chance that the sold capacity or more is available. This is when the prognosis is assumed to be correct.
    The probability is then also 10\%, that the entire sold capacity is not available. If this were to happen, it does not entail that the sold capacity is not available at all, however just that a part of the total capacity is not available. The available part will with a high probability be close to the sold capacity.''}
\end{definition}

Definition \ref{def:P90} allows flexibility providers, including EV aggregators, to place bids in the FCR-D up and down markets during the day-ahead stage, provided that the \textit{probability} of successfully realizing the bid is at least $90$\%. This probability must be based on the aggregator's forecast of its future consumption baseline for the day ahead, and the method used to generate this forecast must be \textit{pre-qualified} by Energinet. Once the prognosis method is verified \textit{ex-ante} by Energinet, the service provider becomes eligible to participate in the ancillary service markets.

Additionally, Energinet conducts an \textit{ex-post} evaluation to monitor how often a service provider's bid is not fully available. It is important to note that unavailability does not necessarily imply a failed response to an activation event. Instead, Energinet checks the \textit{realized} consumption level, regardless of whether an activation event occurred, and determines whether the reserve bid was available as promised. If a bid is partially or fully unavailable, this results in a \textit{reserve shortfall}, also known as \textit{overbidding}.

Energinet typically conducts these ex-post checks in $3$-month periods. During this time, it assesses how often reserve shortfalls occur, and if the frequency violates the $\rm{P90}$ requirement, the corresponding service provider will be excluded from the market. The provider will then need to re-apply for pre-qualification by Energinet. 
In practice, Energinet allows minor errors, which result in the flexibility not being fully available at least 90\% of the time over the past three months. However, significant negative deviations from this threshold are not permitted \cite{energinet2023prequalification}.

Although Definition \ref{def:P90} does not explicitly limit the \textit{magnitude} of reserve shortfalls, it does imply that such shortfalls should not be extreme, thereby discouraging severe overbidding.

\subsection{The $\rm{LER}$ requirement} 
For conventional flexible resources bidding in the FCR-D up and down markets, a full activation must be possible for at least two hours continuously. However, for energy-limited resources, referred to as limited energy reservoirs (LER) by Energinet, such as batteries and EV aggregators, this two-hour requirement has been relaxed to a minimum of just $20$ minutes. In turn, Energinet imposes additional constraints, known as the $\rm{LER}$ requirement, specifically for these types of units. The details of this requirement are as follows. Again, the text is borrowed from \cite{energinet2023prequalification}.
\begin{definition}[The $\rm{LER}$ requirement]\label{def:LERR}
    \textit{[...] ``For FCR-D you must reserve 20\% of the prequalified FCR-D amount to [Normal State Energy Management] NEM in the opposite direction. E.g., if you wish to prequalify 1 MW for FCR-D upwards, you must reserve 0.2 MW in the downwards direction for NEM as well as 20 minutes of full FCR-D upwards delivery, or 0.33 MWh of energy.''}
\end{definition}

The interpretation of this definition is that for an LER unit, bidding $1$ MW in a certain direction of FCR-D requires a \textit{buffer} of $0.2$ MW in to ensure the unit can also provide reserve in the opposite direction. Therefore, a $1$-MW bid requires a $1.2$-MW LER unit. For EV aggregators, we apply the LER requirement only in the downwards direction, since activation (i.e., increasing power consumption) is limited by the battery's energy storage capacity (in MWh). In the upwards direction, when activated, the EV aggregator simply reduces its consumption, similar to other demand-side non-LER flexible resources.

\subsection{Bidding model of the EV aggregator}
Following the $\rm{P90}$ requirement as per Definition \ref{def:P90} and the  $\rm{LER}$ requirement as per
Definition \ref{def:LERR}, we now develop a joint chance-constrained optimization program for the bidding decision-making problem of the EV aggregator. In Denmark, reserve providers submit their FCR-D capacity bids with hourly granularity\footnote{There are currently discussions regarding a change in the bidding resolution to $15$ minutes.}. Therefore, the EV aggregator solves its bidding problem on an hourly basis to determine its FCR-D up $c_{h}^{\uparrow}$ and FCR-D down $c_{h}^{\downarrow}$ capacity bids. This problem for hour $h$ is formulated as follows:
\begin{subequations}\label{General-OPT}
    \begin{align}
        & \underset{(c_{h}^{\downarrow}, c_{h}^{\uparrow}) \geq 0 }{\operatorname{Maximize}}  \quad c_{h}^{\downarrow} + c_{h}^{\uparrow} \label{aab1}\\
        & \text{s.t.} \ \ \ \mathbb{P}\left( \begin{array}{@{}l@{}} 
        \frac{1}{5} c_{h}^{\downarrow} + c_{h}^{\uparrow} \leq F_{m,h}^{\uparrow}, \quad \forall m \\
        c_{h}^{\downarrow} \leq F_{m,h}^{\downarrow}, \quad \forall m \\
        c_{h}^{\downarrow} \leq F_{m,h}^{\mathrm{E}}, \quad \forall m
        \end{array} \right) \geq 1-\epsilon, \label{aab2}
    \end{align}
\end{subequations}
where $m=\{1,...,60\}$ represents the minutes within hour $h$, and $\{F_{m,h}^{\uparrow}, F_{m,h}^{\downarrow}, F_{m,h}^{\mathrm{E}}\}$ are three minute-level random variables that characterize the available flexibility. A bid in hour $h$ is considered fully available if all the minute-level constraints are satisfied.

The objective function \eqref{aab1} maximizes the total quantity of FCR-D up and down bids. If the aggregator has access to forecasted reservation prices, or if it prioritizes one of the two services, it is straightforward to incorporate these prices or priorities as weights in \eqref{aab1}. For the sake of generality, we assume equal weights (i.e., one) for both directions.

The joint chance constraint \eqref{aab2} is formulated according to the $\rm{P90}$ and $\rm{LER}$ requirements. The term $\mathbb{P}\left(.\right) \geq 1-\epsilon$ ensures that the probability of meeting all minute-level probabilistic constraints within the hour is at least $1-\epsilon$, with $\epsilon=0.1$ as specified by the $\rm{P90}$ requirement.

In accordance with the $\rm{LER}$ requirement, the first constraint of \eqref{aab2}  limits the FCR-D down bid $c_{h}^{\downarrow}$  based on the available FCR-D up flexibility $F_{m,h}^{\uparrow}$ (in MW). This constraint does not necessarily require both capacity bids $c_{h}^{\downarrow}$ and $c_{h}^{\uparrow}$ to be non-zero simultaneously. However, if  $c_{h}^{\downarrow}$ is non-zero, it will be restricted not only by the available FCR-D down flexibility $F_{m,h}^{\downarrow}$ (in MW) as stated in the second constraint of \eqref{aab2}, but also by the available FCR-D up flexibility $F_{m,h}^{\uparrow}$, as imposed by the first constraint of \eqref{aab2}. Note that a constraint such as $c_{h}^{\uparrow} \leq F_{m,h}^{\uparrow}$ would be redundant, as the first constraint of \eqref{aab2} is more binding. Further, according to Definition \ref{def:LERR}, the FCR-D down bid $c_{h}^{\downarrow}$ must also be limited by the capability of the LER unit to be fully activated for at least $20$ minutes continuously. This requirement is captured by the last constraint of \eqref{aab2}, where $F_{m,h}^{\mathrm{E}}$ represents the maximum power (in MW) that the EV aggregator can add to the fleet's consumption for the next $20$ minutes, without exceeding the energy storage capacity (in MWh) of the EVs, subject to their maximum charging power.

\begin{figure*}[t]
    \centering
    \includegraphics[width=\textwidth]{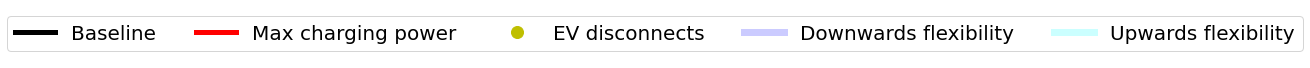}
    \begin{subfigure}{.320\textwidth}
        \vspace{-2mm}
        \centering
        \includegraphics[width=\columnwidth]{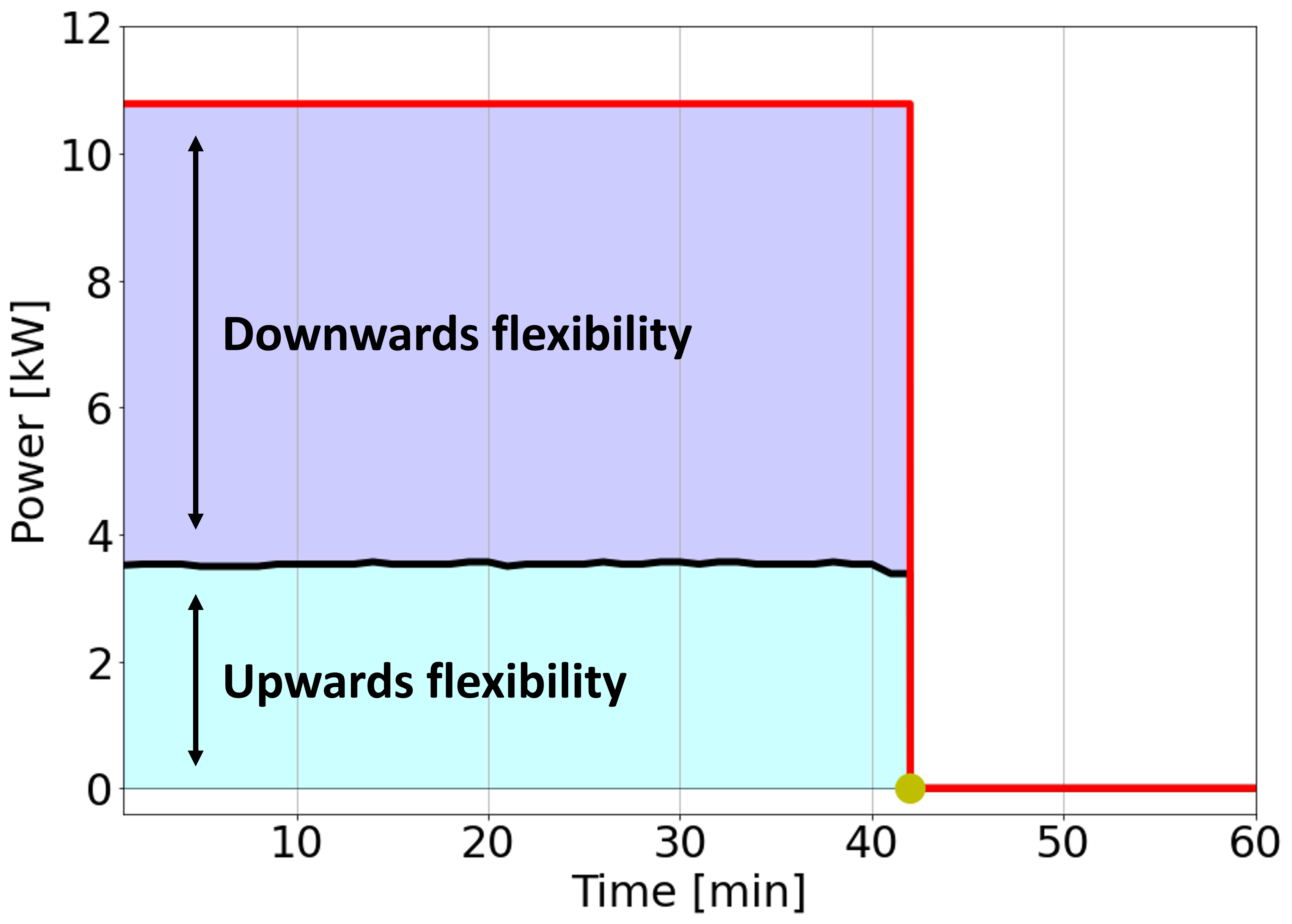}
        \caption{One EV without the $\rm{LER}$ requirement}
        \label{fig:1CBwoLER}
    \end{subfigure}%
    \hfill
    \begin{subfigure}{.339\textwidth}
        \vspace{-2mm}
        \centering
        \includegraphics[width=\columnwidth]{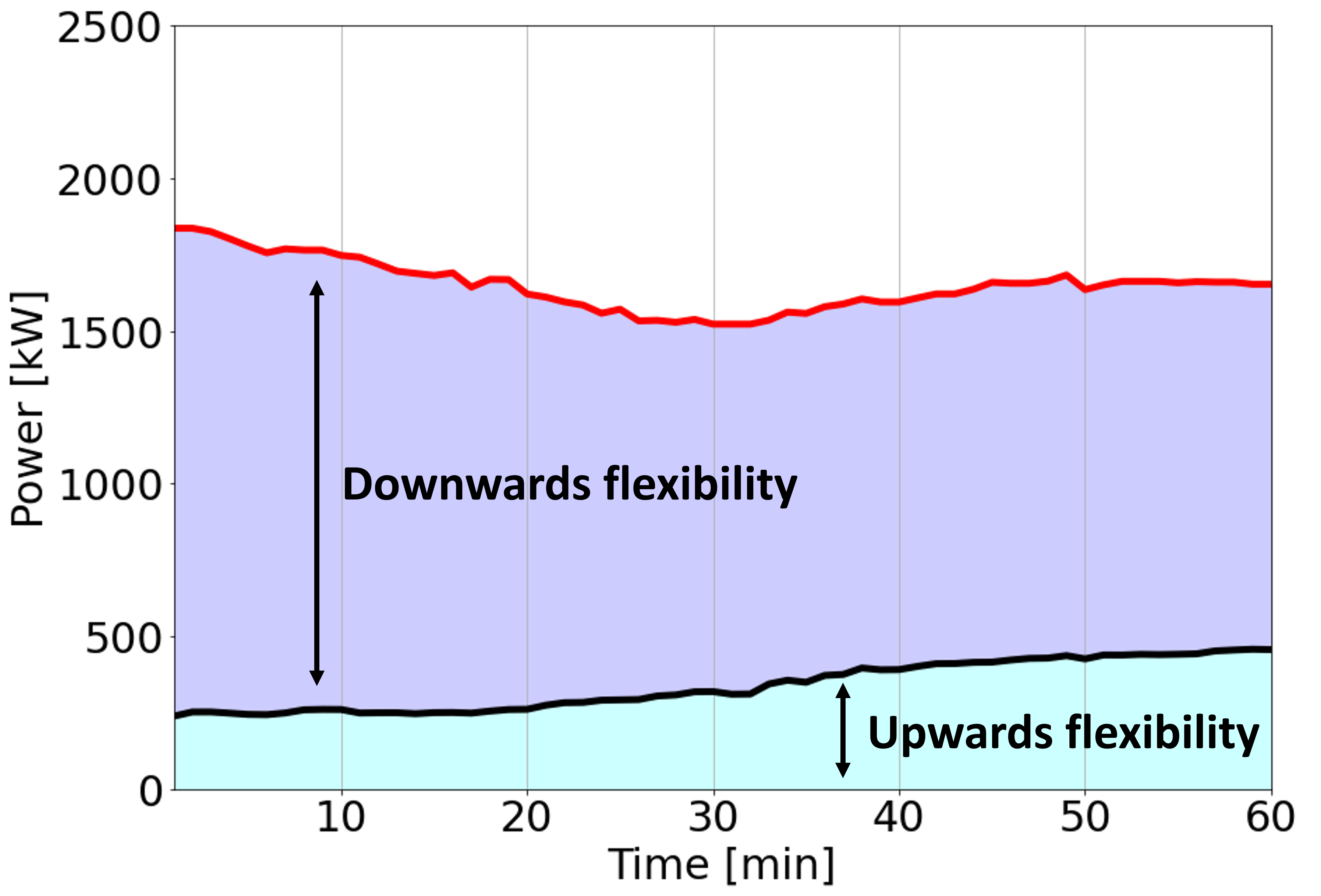}
        \caption{$1400$ EVs without the $\rm{LER}$ requirement}
        \label{fig:1400CBwoLER}
    \end{subfigure}%
    \hfill
    \begin{subfigure}{.339\textwidth}
        \vspace{-2mm}
        \centering
        \includegraphics[width=\columnwidth]{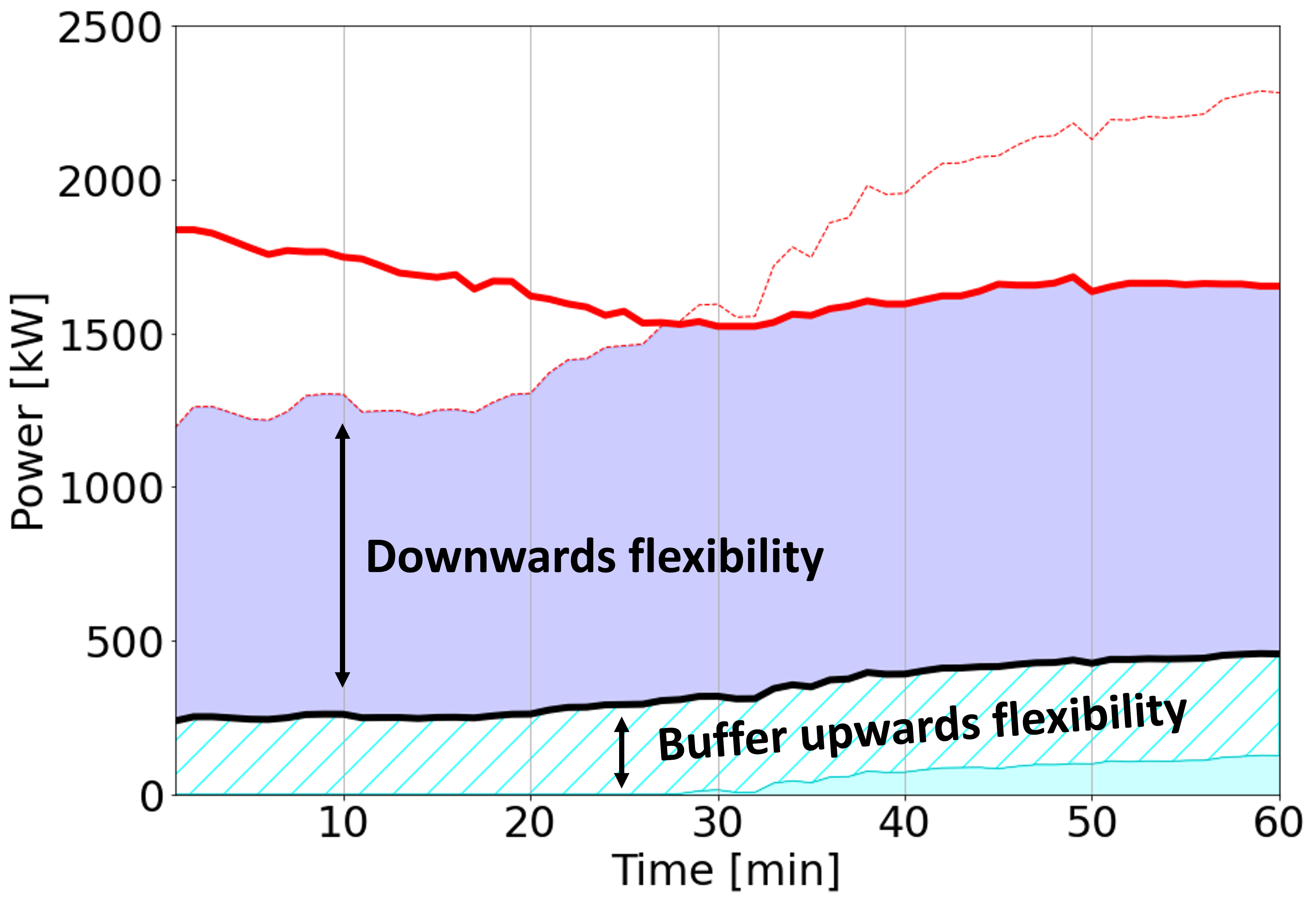}
        \caption{$1400$ EVs with the $\rm{LER}$ requirement}
        \label{fig:1400CBwithLER}
    \end{subfigure}
    \caption{\small{The historical consumption level (baseline) and the available capacity for upwards and downwards flexibility in a random hour for one or the aggregation of $1400$ EVs with and without the implementation of the $\rm{LER}$ requirement.}}
    \label{fig:assets_ind}
\end{figure*}


Chance-constrained programs are typically computationally intractable. If the underlying probability distribution exhibits certain properties, an analytical reformulation of the chance constraints can be developed \cite{shapiro}. To maintain generality and avoid assuming a specific distribution for our empirical data, we apply sample-based techniques to reformulate the chance constraints. Exploring the potential benefits of fitting a specific distribution to the data, rather than relying on the empirical distribution, is left for future work. Through sampling, the random variables $F^{\uparrow}_{m,h}$, $F^{\downarrow}_{m,h}$, $F^{\rm{E}}_{m,h}$ are represented by the set of samples $F^{\uparrow}_{m,h,\omega}$, $F^{\downarrow}_{m,h,\omega}$, $F^{\rm{E}}_{m,h,\omega}$. The next section outlines how these samples are obtained from historical data.

\section{Flexibility estimation and drawing samples}\label{sec:est}

This section begins by illustrating the estimation of available flexibility for a random hour. We then proceed to describe the probabilistic approach used for estimating available flexibility. Following this, we generate samples for both in-sample and out-of-sample analyses. Finally, we outline the setup employed for cross-validation.

\subsection{Estimation of available flexibility for a single random hour}

\begin{figure*}[t]
    \centering
    \begin{subfigure}{.315\textwidth}
        \centering
        \raisebox{0.00cm}{ \includegraphics[height=3.65cm]{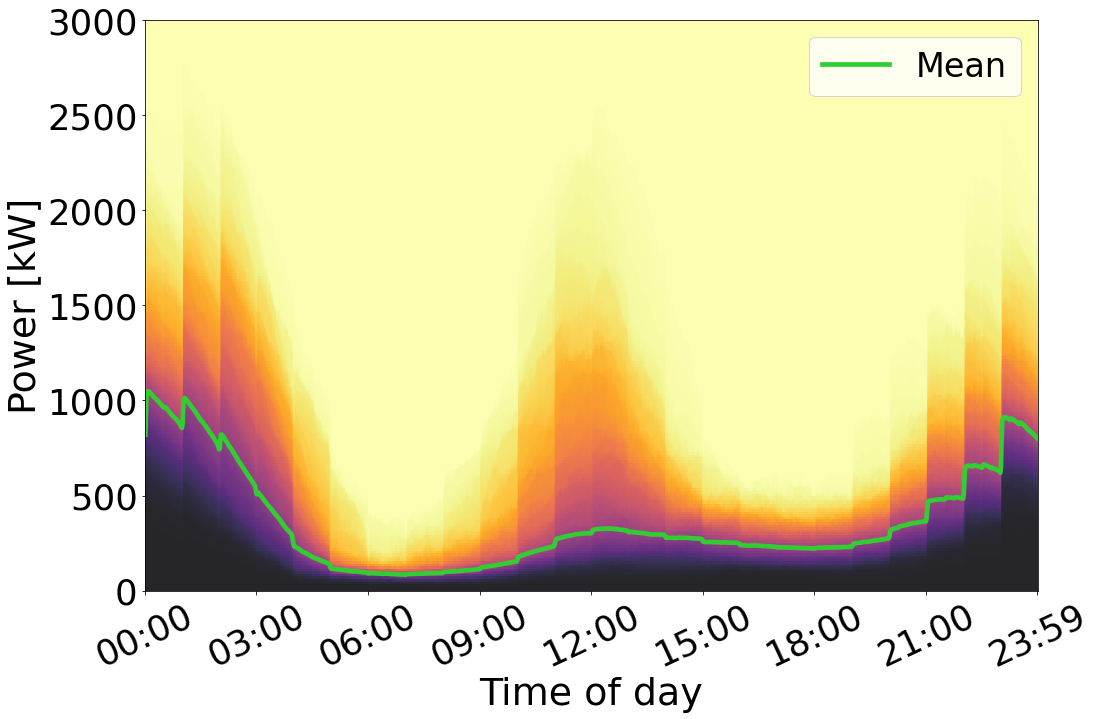}}
        \caption{Upwards flexibility $F_{m,h}^{\uparrow}$}
        \label{fig:upwards}
    \end{subfigure}%
    \hfill
    \begin{subfigure}{.315\textwidth}
        \centering
        \raisebox{0.0cm}{ \includegraphics[height=3.65cm]{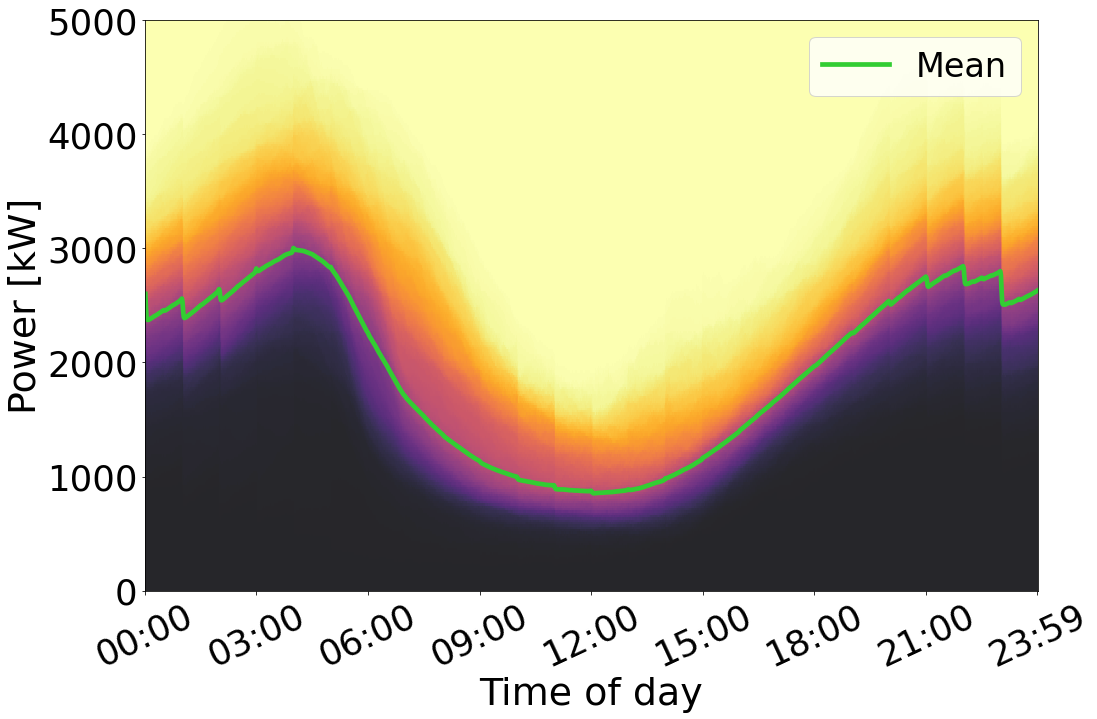}}
        \caption{Downwards flexibility $F_{m,h}^{\downarrow}$}
        \label{fig:downwards}
    \end{subfigure}%
    \hfill
    \begin{subfigure}{.315\textwidth}
        \centering
        \includegraphics[height=3.64cm]{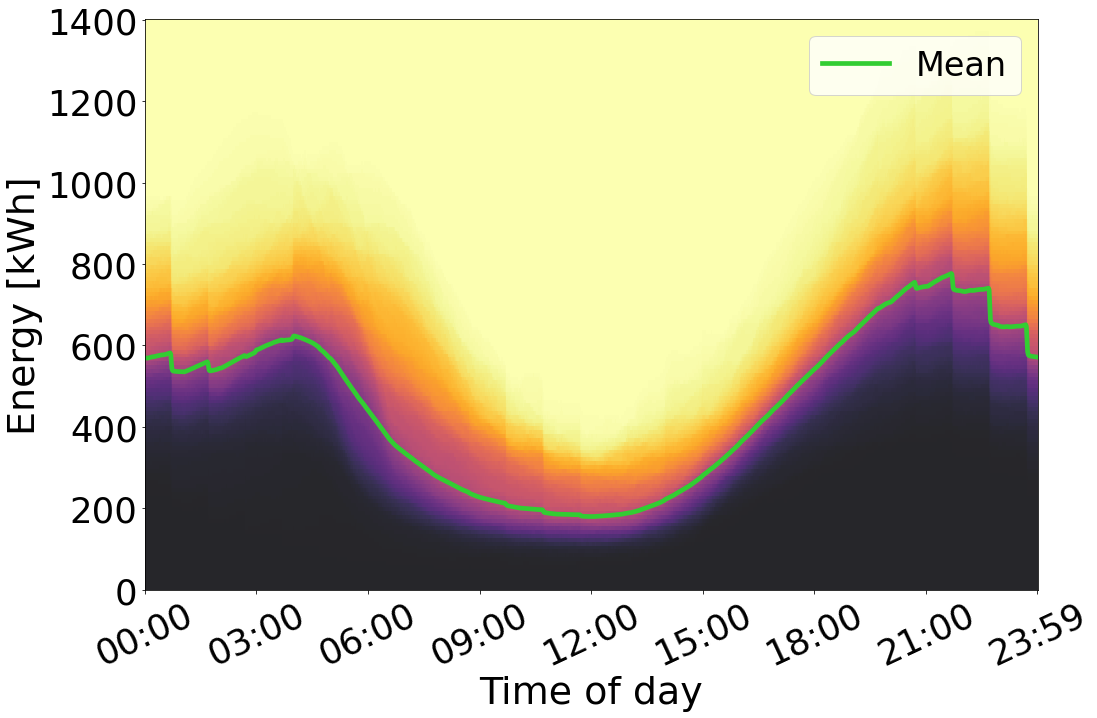}
        \caption{Energy flexibility $\frac{F_{m,h}^{\mathrm{E}}}{3}$}
        \label{fig:energy}
    \end{subfigure}
    \hfill
    \begin{subfigure}{.04\textwidth}
        \centering
        \raisebox{1.095cm}{\includegraphics[height=3.13cm]{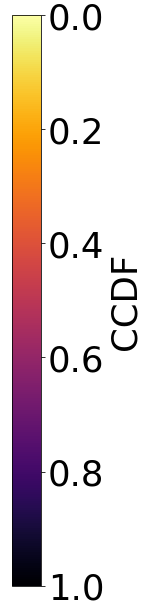}} 
    \end{subfigure}%
    \caption{\small{Conditional cumulative distribution function (CCDF) of available flexibility for a portfolio of $1400$ EVs from March $24$, $2022$, to March $21$, $2023$.}}
    \label{fig:flexibility}
    \vspace{-1mm}
\end{figure*}

This study leverages real-world data collected from $1400$ EV charging stations located at residential properties across Denmark. The dataset spans from March $24$, $2022$, to March $21$, $2023$, capturing detailed charging activity over this period. It is assumed that each charging station is dedicated to a single EV, ensuring consistency in the data. On average, measurements were registered every $2.84$ minutes, and from this, minute-resolution consumption profiles have been interpolated for each EV. The historical EV consumption level serves as the baseline for flexibility estimation.

Fig. \ref{fig:assets_ind}  illustrates the historical consumption level (baseline) for two different EV portfolio sizes at a random hour. Fig. \ref{fig:assets_ind}(a) shows the consumption baseline of a single EV, which is disconnected from the grid at minute $42$. While connected, the upwards flexibility refers to its ability to reduce consumption to zero, while the downwards flexibility refers to the potential to increase consumption up to its maximum charging power (in MW). Note that the battery energy storage capacity constraint (in MWh), included in the $\rm{LER}$ requirement, is not enforced in this plot. 

Fig. \ref{fig:assets_ind}(b) depicts a similar scenario but for a portfolio of $1400$ EVs.  Since the number of grid-connected EVs changes throughout the hour, the maximum charging power (represented by the thick red curve) is time-variant. 

Fig. \ref{fig:assets_ind}(c) is similar to Fig. \ref{fig:assets_ind}(b) but incorporates the $\rm{LER}$ requirement. To facilitate interpretation, let us assume the aggregator prioritizes bidding in the FCR-D down market over the FCR-D up market due to its comparatively higher historical prices (this assumption is for illustration purposes; the model does not include this assumption). The first and third constraints in \eqref{aab2} are key to understanding the feasibility of reserve capacity bids. In this example, the first constraint is binding, while the third is not. Let us explain how the first constraint limits the feasible space for reserve capacity bids.

At the first minute in Fig. \ref{fig:assets_ind}(c), the available upwards flexibility (equal to the baseline) is approximately $F_{m,h}^{\uparrow}=200$ kW. In the same minute, the available downwards flexibility is about $F_{m,h}^{\downarrow}=1600$ kW, which equals the total capacity of $1800$ kW (represented by the thick red curve) minus the consumption level of $200$ kW.  According to the first constraint, the \textit{maximum} FCR-D down bid that the aggregator can place is $5F_{m,h}^{\uparrow}$, i.e., $1000$ kW, assuming that $c_{h}^{\uparrow}=0$. Therefore, in minute $1$, the maximum available downwards flexibility is $1000$ kW (depicted by the thin red curve), rather than the full $1600$ kW. The shaded blue area under the consumption baseline indicates that although there is potential for upwards flexibility, it is kept as a buffer to allow for the sale of downwards flexibility. 
If $c_{h}^{\uparrow}$ were allowed to take a non-zero value, it would further restrict the upper bound for $c_{h}^{\downarrow}$. After minute $30$, the first constraint is no longer binding on $c_{h}^{\downarrow}$, as long as the FCR-D up bid ($c_{h}^{\uparrow}$) remains low enough. As a result, the aggregator can place non-zero bids for both FCR-D up and down services simultaneously.

We now focus on the $\rm{LER}$ requirement enforced by the third constraint in \eqref{aab2}. Recall that $F_{m,h}^{\mathrm{E}}$ represents the maximum amount of energy that a given EV can consume in the upcoming $20$ minutes. To calculate $F_{m,h}^{\mathrm{E}}$, we use historical data on each EV's charging session to estimate the initial state of charge when the vehicle first connects to the grid. For the end of the charging session, we assume the EV reaches $90$\% state of charge, as this is commonly recommended by manufacturers to mitigate battery degradation issues \cite{MCALEER}. 

Even when an EV is connected to the grid but not actively charging, it can still provide FCR-D down services using the remaining $10$\% of its energy storage capacity. Due to data limitation, we define the excess battery capacity available during each chargeging session as the $10$\% of its largest charging session (kWh) in the dataset.

Additionally, the EVs will be physically constrained by their maximum charging power, meaning $F^{\downarrow}_{m,h}$ serves as an upper bound for $F_{m,h}^{\mathrm{E}}$. This method provides a rather conservative estimate of the energy flexibility, ensuring that the second constraint in \eqref{aab2} will never become binding, as $F_{m,h}^{\mathrm{E}}$ will always be equal to or lower than $F^{\downarrow}_{m,h}$.

\maybe{\textcolor{red}{\sout{In the context of our real-world case study, we found that the third constraint in \eqref{aab2} does not bind. However, it is important to note that this may not hold for other case studies, particularly those with different configurations or datasets, where the third constraint may be more restrictive.}}}

Finally, we emphasize that Fig. \ref{fig:assets_ind}(c) is provided solely for illustrative purposes to clarify how the $\rm{LER}$ requirement may impact the bidding problem. In practice, the input data for available upwards flexibility ($F_{m.h}^{\uparrow}$) and downwards flexibility ($F_{m,h}^{\downarrow}$) remain as shown in Fig. \ref{fig:assets_ind}(b). In other words, we do not modify $F_{m,h}^{\uparrow}$ and $F_{m,h}^{\downarrow}$ ex-ante based on the $\rm{LER}$ requirement. Instead, it is the optimization problem itself that determines the extent to which the $\rm{LER}$ requirement constrains the optimal bidding decisions.
Recall that the optimal bids for upwards ($c_{h}^{\uparrow}$) and downwards flexibility ($c_{h}^{\downarrow}$) are fixed over the duration of an hour, while $F_{m,h}^{\uparrow}$ and $F_{m,h}^{\downarrow}$ can vary throughout that hour.

\vspace{4mm}

\subsection{Probabilistic estimation of the available flexibility}

Instead of focusing on a single random historical hour, we now expand our analysis to consider all historical data in order to derive a probabilistic estimation of the available flexibility. This estimation will serve as a forecast for the available flexibility during each hour of the following day. In other words, rather than using historical data to probabilistically predict future flexibility, we treat the historical data as a proxy for the forecast.
For simplicity, we do not employ any classification methods or advanced techniques to account for seasonality or other external factors. Instead, we utilize \textit{all} available historical data to estimate flexibility for the upcoming day. A potential avenue for future research would be to introduce conditionality—specifically, selecting the most relevant historical samples based on current operating conditions, which could improve the accuracy of the flexibility forecasts.

Using all historical samples, Fig. \ref{fig:flexibility} illustrates the distribution of various types of flexibility, including upwards ($F_{m,h}^{\uparrow}$), downwards ($F_{m,h}^{\downarrow}$), and energy ($F_{m,h}^{\rm{E}}/3$), at a minute-level resolution throughout the day for a portfolio containing all $1400$ EVs. The final plot in the figure shows the maximum additional amount of energy (in kWh) that can be consumed in the next $20$ minutes without exceeding the energy storage constraint of the portfolio. This value can be multiplied by $3$ to convert it to $F_{m,h}^{\rm{E}}$ (in kW), as shown in \eqref{aab2}. The conditional cumulative distribution function (CCDF) provides the probability associated with the EV aggregator having a certain level of flexibility at any given time. For example, based on Fig. \ref{fig:flexibility}(a), the probability that the available upwards flexibility in the first minute will be $1000$ kW or less is around $60$\%, while there is a $100$\% probability that it will be $500$ kW or less. We observe that the distribution of upwards flexibility shows a larger spread compared to the downwards flexibility and energy flexibility distributions. Notably, in the middle of the day, there is a significant right-tailed skew in the upwards flexibility, suggesting that the flexibility is more likely to increase significantly than to decrease. Furthermore, there is a notable discrepancy between the available upwards and downwards flexibility. This indicates that, in this case study, the primary source of flexibility for the EV fleet is their ability to increase consumption, rather than reduce it. For readers interested in exploring more robust approaches to flexibility estimation, we refer to \cite{karan}, which discusses distributionally robust methods for estimating flexibility in EV aggregators, particularly useful in cases where data availability is limited.

\subsection{Scenario generation for the in-sample analysis} \label{Scenario}

Here, we describe the process of drawing a set of samples (also referred to as scenarios) that will be used when reformulating the joint chance-constrained program \eqref{General-OPT}. This set of samples, denoted as $\omega\!\in\!\Omega$, is referred to as
\textit{in-sample} scenarios, and the proposed chance-constrained optimization model serves as the foundation for our in-sample analysis. The sample set should capture the uncertain nature of three random variables: available upwards flexibility, downwards flexibility, and energy flexibility, which are grouped as $\{F^{\uparrow}_{m,h,\omega}, F^{\downarrow}_{m,h,\omega}, F^{\text{E}}_{m,h,\omega}\}$ for each minute $m$ of hour $h$. These uncertain parameters are already depicted in Fig.     \ref{fig:flexibility}. 

It is important to note that, while we represent the available flexibility and its corresponding uncertainty at the minute level, the reserve capacity bids are set on an hourly basis. This means that the bid remains fixed throughout each hour. We randomly draw $|\Omega|$ samples, with each sample corresponding to a real historical profile, thus preserving the potential correlations of flexibility between minutes.

A key consideration is the number of samples, $|\Omega|$, which must be large enough to adequately represent the underlying uncertainty. Although the model will also be evaluated out-of-sample, we follow the analytical findings of \cite{Needed_samples}, which suggest a lower bound for $|\Omega|$ to ensure sufficient representation of uncertainty:
\begin{equation}\label{eq:N_smaples}
    |\Omega| \geq \frac{2}{\epsilon} \log \left(\frac{1}{\delta}\right)+2 n+\frac{2 n}{\epsilon} \log \left(\frac{2}{\epsilon}\right),
\end{equation}
where $\epsilon = 0.1$ corresponds to the $\rm{P90}$ requirement, $n = 2$ represents the number of variables (i.e., $c_{h}^{\downarrow}$ and $c_{h}^{\uparrow}$ for the optimization problem in hour $h$), and $\delta = 0.01$ is our chosen value, resulting in a confidence level of $99$\% that the sample set accurately represents the underlying distribution. This leads to the conclusion that \textit{at least} $216$ randomly-selected samples are required.

It is important to note that \eqref{eq:N_smaples} holds under the assumption that the underlying data-generating process is stationary, which may not always be the case in real-world scenarios. However, our ex-post cross-validation process confirms that the sampling procedure works satisfactorily for our application, even if the data-generating process is not strictly stationary.

\subsection{Ex-post out-of-sample analysis}

We have merged the second and third stages of Fig. 
\ref{fig:timeline_FCR}, where the true consumption of the EV aggregator is realized. Energinet conducts an ex-post evaluation of the occurrence frequency of reserve shortfalls over a given time period, typically the past three months. It is important to note that Energinet performs this check based on reserve availability, regardless of whether the reserve service was actually activated.  If the occurrence of reserve shortfalls exceeds $10$\%\maybe{\textcolor{red}{\sout{(or $15$\% with an additional buffer)}}} of the total time periods, the aggregator loses its qualification to bid in the Nordic ancillary service markets. Conversely, if the ex-post check reveals that the aggregator's performance is satisfactory, it will not lose its qualification. However, the aggregator will still face penalties for unsuccessful activations (i.e., when activation events were initiated but not successfully carried out), while maintaining its eligibility to participate in future bidding processes.

As part of our ex-post out-of-sample analysis, for each realization $\omega^\prime$ (which may differ from the in-sample scenarios $\omega \in \Omega$), we calculate the highest FCR-D up and down shortfall (in kW) within each hour $h$ when the service was activated ($p_{h,\omega^\prime}^{\uparrow}$ and $p_{h,\omega^\prime}^{\downarrow}$). We assess the need for activation based on the historical frequency records obtained from \cite{fingrid_frequency_historical_data_2023}, which provide data on the actual frequency of reserve activation events. If the service was activated but the reserve shortfall was not met, the aggregator is penalized at rates $\lambda_{h}^{\uparrow}$ and $\lambda_{h}^{\downarrow}$ for unsuccessful activations. This penalty is applied separately for both upwards and downwards reserve shortfalls, reflecting the cost of not fulfilling the reserve obligations when required.
Energinet defines these penalty rates as the substitution cost for replacing a missed activation. However, since no historical data on the replacement cost exists, we assume a value equal to the corresponding reservation price multiplied by five, i.e., $\lambda_{h}^{\uparrow}=5\pi_{h}^{\uparrow}$ and $\lambda_{h}^{\downarrow}=5\pi_{h}^{\downarrow}$. Thus, the eventual out-of-sample profit for the aggregator under realization $\omega^\prime$ is given by:

\begin{align}
    \underbrace{\sum_h \Big(c_{h}^{\uparrow} \pi_{h}^{\uparrow} + c_{h}^{\downarrow} \pi_{h}^{\downarrow}\Big)}_{\text{Reservation payment (in-sample)}} - \underbrace{\sum_h \Big(\lambda_{h}^{\uparrow} p_{h,\omega^\prime}^{\uparrow} + \lambda_{h}^{\downarrow} p_{h,\omega^\prime}^{\downarrow}\Big)}_{\text{Penalty cost (out-of-sample)}}. \label{eq:obj}
\end{align}

\subsection{Cross validation} \label{cross}

Given access to historical data for $363$ days and the requirement for at least $216$ samples for our in-sample analysis, we conducted a $3$-fold cross-validation simulation. For each hour $h\!=\!\{1,2,...,24\}$, we randomly divide the available historical data (i.e., $363$ samples) into three groups. Two of these groups, containing a total of $242$ samples (which satisfies the requirement for at least $216$ samples), are used for the in-sample analysis, while the remaining group, containing $121$ samples, is used for the out-of-sample analysis.
We repeat this procedure three times, each time using a different combination of groups for the in-sample and out-of-sample analyses. This approach ensures that each subset of historical data is used for both in-sample and out-of-sample analyses, preventing any bias that might arise from the random selection of scenarios.
The final results are reported as the \textit{average} over the three simulations, allowing us to evaluate the model's performance across different scenario groupings and ensure robustness in the findings.

\vspace{-2mm}
\section{Sample-based reformulation of \eqref{General-OPT}}\label{sec:math-modelling}

Given the in-sample scenarios generated in the previous section, we now present two sample-based reformulations of the joint chance-constrained problem \eqref{General-OPT}, along with a heuristic approach, referred to as the naive approach, which does not require solving any optimization problem.

\vspace{3mm}
\subsection{ALSO-X algorithm}

Problem \eqref{General-OPT} for every hour $h$ can be reformulated  as
%
%
\begin{subequations}\label{ALSO-MILP}
    \begin{equation}
        \label{c1_a}
        \underset{(c_{h}^{\downarrow}, c_{h}^{\uparrow}) \geq 0, \ y_{m,h,\omega} \in \{0,1\}}{\operatorname{Maximize}} \ \ \ \ c_{h}^{\downarrow}+c_{h}^{\uparrow}
    \end{equation}
\vspace{-2mm}
    \begin{align}
        \label{q1_a}
         \text{s.t.} \ \ \  \frac{1}{5} c_{h}^{\downarrow}+c_{h}^{\uparrow}-F^{\uparrow}_{m,h,\omega} & \leq y_{m,h,\omega}  \ \! M^{\uparrow}   & \forall m, \omega \\
        \label{q2_a}
        c_{h}^{\downarrow}-F^{\downarrow}_{m,h,\omega}                        & \leq y_{m,h,\omega}  \ \!  M^{\downarrow} & \forall m, \omega \\
        \label{q5_a}
        c_{h}^{\downarrow}-F^{\rm{E}}_{m,h,\omega}                        & \leq y_{m,h,\omega} \ \!M^{\rm{E}}          & \forall m, \omega \\
        \sum_{m,\omega} y_{m,h,\omega}                           & \leq Q, \label{acx}
    \end{align}
\end{subequations}
where $\{M^{\uparrow}, M^{\downarrow}, M^{\rm{E}}\}$ are large enough positive constants, and $y_{m,h,\omega}$ is an auxiliary binary variable corresponding to minute $m$, hour $h$, and sample $\omega$.  This makes \eqref{ALSO-MILP} a mixed-integer linear program. If $y_{m,h,\omega}\!=\!0$, it implies that all three minute-level constraints within \eqref{aab2} are satisfied for sample $\omega$. On the other hand, if $y_{m,h,\omega}\!=\!1$, at least one of those three minute-level constraints is violated, and the big-$M$ values in \eqref{q1_a}-\eqref{q5_a} ensure that the optimization problem remains feasible. Constraint \eqref{acx} enforces the budget $Q$, which limits how many constraint violations are allowed across the minutes and samples. We set the budget $Q$ to $\epsilon = 0.1$ times $60 |\Omega|$. Additionally, the values of ${M^{\uparrow}, M^{\downarrow}, M^{\rm{E}}}$ are set to the largest available value of corresponding flexibility from our empirical data.

One potential computational challenge when solving \eqref{ALSO-MILP} is the large number of binary variables, one for each minute and sample. To address this, \cite{Ahmed201751} introduces the ALSO-X algorithm, which finds a feasible solution to \eqref{ALSO-MILP} by iteratively solving its linear relaxation, where the binary variables are relaxed to continuous values between 0 and 1, i.e., $0 \leq y_{m,h,\omega}\!\leq\!1$. This approach is detailed in Algorithm \ref{alg:also-x}.  For our implementation, we set $\delta$ to  $10^{-5}$.




\begin{algorithm}[t]
    \caption{ALSO-X}\label{alg:also-x}
    \begin{algorithmic}
        \Statex \text{Input:} Stopping tolerance parameter $\delta$
        \Statex \textbf{Require:} $\text{Relax binary variables} \ y_{m,h,\omega}$
        \State $\underline{Q} \leftarrow 0, \quad \bar{Q}  \leftarrow 60 \epsilon |\Omega|$
        \While{$\bar{Q} - \underline{Q} \geq \delta$}
        \State $\text {Set } Q = \frac{(\underline{Q}+ \bar{Q})}{2}$
        \State $\text {Retrieve } \Theta^* \text { as an optimal solution to relaxed (\ref{ALSO-MILP})}$
        \State $\text {Set } \underline{Q}= Q\text{ if } \mathbb{P}(.) \geq 1-\epsilon \text{; otherwise, } \bar{Q}=Q$
        \EndWhile
        \Statex \text{Output: A feasible solution to  \eqref{ALSO-MILP} }
    \end{algorithmic}
\end{algorithm}


\subsection{CVaR approximation}

Here, we use the CVaR to approximate the joint chance constraint \eqref{aab2} by controlling the \textit{magnitude} of reserve shortfall. The advantage of this approach is that it transforms the optimization model into a linear form. However, the limitation is that it constrains the magnitude of the violation, while the original problem \eqref{General-OPT} does not explicitly aim to do so. As a result, this approximation may lead to a more conservative solution, which the aggregator might find unnecessary. Specifically, this approach limits the expected violation for the worst $(1-\alpha)$ fraction of the samples, where $\alpha = 0.1$, as per
Definition \ref{def:P90}. The resulting linear problem for hour $h$ can be written as:
%
%
\begin{subequations}\label{CVAR-MILP_1}
    \begin{equation}
        \label{cvar1_a_1}
        \underset{(c_{h}^{\downarrow}, c_{h}^{\uparrow}) \geq 0, \ \beta_{h} \leq 0, \ \zeta_{m,h,\omega}}{\operatorname{Maximize}}   \ \ \ \ c_{h}^{\downarrow}+c_{h}^{\uparrow}
    \end{equation}
\vspace{-2mm}
    \begin{align}
        \label{cvar2_a_1}
        \text{s.t.} \ \ \  \frac{1}{5} c_{h}^{\downarrow}+c_{h}^{\uparrow}-F^{\uparrow}_{m,h,\omega}     & \leq \zeta_{m,h,\omega} & \forall m, \omega \\
        \label{cvar3_a_1}
        c_{h}^{\downarrow}-F^{\downarrow}_{m,h,\omega}                            & \leq \zeta_{m,h,\omega} & \forall m, \omega \\
        \label{cvar4_a_1}
        c_{h}^{\downarrow}-F^{\rm{E}}_{m,h,\omega}                            & \leq \zeta_{m,h,\omega} & \forall m, \omega \\
        \label{cvar5_a_1}
        \frac{1}{60|\Omega|} \sum_{m,\omega} \zeta_{m,h,\omega}  & \leq (1-\alpha) \beta_{h}              &                   \\
        \label{cvar6_a_1}
        \beta_{h}                                                               & \leq \zeta_{m,h,\omega} & \forall m, \omega,
    \end{align}
\end{subequations}
where $\zeta_{m,h,\omega}$ and $\beta_{h}$ are auxiliary variables. 

We refer the interested reader to \cite{nan}, which provides a comparative discussion on the performance of the CVaR approximation, the ALSO-X algorithm, and its extended version for solving chance-constrained problems, accompanied by several numerical examples. In the next section, we will discuss the performance of these two methods  for our application.

\subsection{Naive approach}

As our heuristic-based baseline model, we introduce a naive bidding approach that does not require solving an optimization problem. This approach, outlined in Algorithm \ref{alg:Naive}, follows a simple rule-based strategy and consists of three main steps. First, we compute the 10th percentile of the underlying flexibility distribution for hour $h$ based on the in-sample scenarios. For instance, to determine this value for upwards flexibility in hour $h$, i.e., $F^{\text{P}\uparrow}_h$, we apply the percentile operator $P_{0.1} \left( F^{\uparrow}_{m,h,\omega} \ \forall{m},\forall{\omega}\right)$, considering all samples $\omega$ and minutes $m$ within that hour. This calculation is performed at the hourly level rather than per minute because FCR-D bids are submitted at an hourly resolution. As a result, we obtain the percentile values $F^{\text{P}\uparrow}_h$, $F^{\text{P}\downarrow}_h$, and $F^{\text{P}\rm{E}}_h$ for hour $h$. Second, in accordance with the $\rm{LER}$ requirements, we set the FCR-D down bid as the minimum of $5F^{\text{P}\uparrow}_h$, $F^{\text{P}\downarrow}_h$, and $F^{\text{P}\rm{E}}_h$. Third, again following the $\rm{LER}$ requirements, we define the FCR-D up bid as the 10th percentile of the upwards flexibility minus one-fifth of the FCR-D down bid. It is important to note that this heuristic approach considers each $\rm{LER}$ requirement independently, which may lead to a higher violation rate.

\begin{algorithm}[t]
    \caption{Naive approach}\label{alg:Naive}
    \begin{algorithmic}
         \Statex \text{Input:} Sampled flexibilities $F^{\uparrow}_{m,h,\omega}$, $F^{\downarrow}_{m,h,\omega}$, and $F^{\rm{E}}_{m,h,\omega}$
        \For{$h$} 
       \State  \ensuremath{ \text{set } 
        F^{\text{P}\uparrow}_h = P_{0.1} \left( F^{\uparrow}_{m,h,\omega} \ \forall{m},\forall{\omega}\right)
        }
\State  \ensuremath{\text{set } 
        F^{\text{P}\downarrow}_h = P_{0.1} \left( F^{\downarrow}_{m,h,\omega} \ \forall{m},\forall{\omega}\right)
        }
\State  \ensuremath{ \text{set } 
        F^{\text{P}\rm{E}}_h = P_{0.1} \left( F^{\rm{E}}_{m,h,\omega} \ \forall{m},\forall{\omega}\right)
        }
        \State \ensuremath{\text{Retrieve } c_{h}^{\downarrow} = \min \left\{ 5 F^{\text{P}\uparrow}_h, F^{\text{P} \downarrow}_h, F^{\text{P}\rm{E}}_h \right\}
        }
        \State \ensuremath{\text{Retrieve } c_{h}^{\uparrow} = F^{\text{P} \uparrow}_h - \frac{1}{5}c_{h}^{\downarrow}}
        \EndFor
    \Statex \text{Output: FCR-D up and down bids} $(c_{h}^{\downarrow}, c_{h}^{\uparrow})$   
    \end{algorithmic}
\end{algorithm}

\vspace{3mm}
\section{Numerical results}\label{sec:results}
This section presents numerical bidding results obtained using the ALSO-X and CVaR methods. For comparison, it also includes results from an oracle model and the naive approach. The oracle model has perfect foresight into the future consumption of the EV portfolio. Although not practical, the oracle model serves as an upper bound for potential profit, offering insight into the extent to which the performance of the proposed chance-constrained models could be improved. Due to its simplicity, the naive method serves as our baseline model for adhering to the $\rm{P90}$ requirement.

We begin by presenting the profit results, followed by an investigation of the synergy effect on the profit and reserve capacity bids placed. Lastly, we explore the frequency of overbidding (reserve shortfall) and assess whether the 
$\rm{P90}$ requirement is met. All source codes for this work are publicly available at \cite{githubrepo}, although the historical EV consumption data remains proprietary.



\subsection{Profit}
Consider the case where all $1400$ EVs are within the portfolio of a single aggregator. Assuming the aggregator fully distributes the profit uniformly among the EVs, each EV in the portfolio would earn an annual profit of $857$ DKK using the ALSO-X method and $638$ DKK with the CVaR method, by bidding $24.4$\% and $18.5$\% of its total flexibility, respectively. This results in a cost saving of $6$-$10$\% for a typical EV in Denmark over the course of a year.

The main takeaway from this analysis is that there is a significant monetary benefit to pooling available EVs and bidding their flexibility in the FCR-D markets.


\subsection{Synergy effect quantification}
We hypothesize that there is a synergy effect when aggregating EVs, such that a larger portfolio of EVs can place a higher reserve quantity bid per EV compared to multiple smaller portfolios with the same total number of EVs. To validate this hypothesis, we define two key metrics.

The first metric is the aggregator's profit, as given by \eqref{eq:obj} which is calculated through the $3$-fold cross-validation process. The second metric is the utilized bidding capacity (UBC), defined as the percentage of total flexibility bid for each hour 
$h$, given by:
\begin{align}
    {\rm{UBC}}_{h}^{\uparrow} = \frac{ c_{h}^{\uparrow}}{\frac{1}{60}\sum_{m} F_{m,h}^{\uparrow}}; \ \ {\rm{UBC}}_{h}^{\downarrow} = \frac{ c_{h}^{\downarrow}}{\frac{1}{60}\sum_{m} F_{m,h}^{\downarrow}} \label{sey1}.
\end{align}

\begin{figure}[t!]
    \centering
    \includegraphics[width=1\columnwidth]{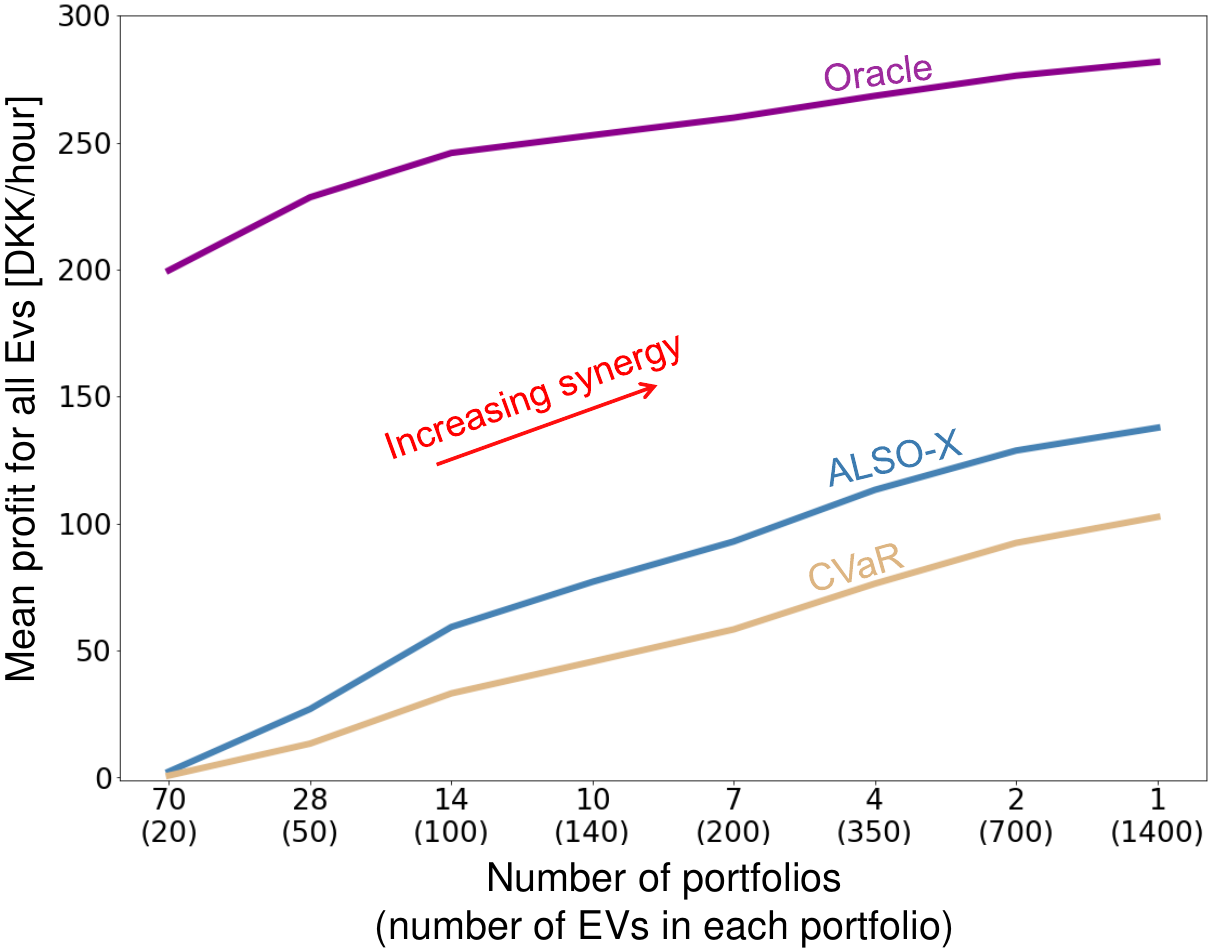}
    \caption{\small{The mean profit (DKK/hour) of the aggregator is shown on the y-axis. On the x-axis, the first case represents the scenario where the $1400$ EVs are divided into $70$ portfolios of $20$ EVs each (i.e., the aggregator bids for each portfolio separately), while the last case corresponds to the largest portfolio where the aggregator bids the flexibility of all $1400$ EVs together. The profit for each method (oracle, ALSO-X, and CVaR) is calculated out-of-sample as per \eqref{eq:obj}, following the $3$-fold cross-validation process described in Section \ref{cross}.}}
\label{fig:Synergy_bundle}
\end{figure}

An increase in both metrics, \eqref{eq:obj} and \eqref{sey1}, would indicate a synergy effect. 
Fig. \ref{fig:Synergy_bundle} shows the mean profit of the aggregator (in DKK/hour) when the  $1400$ EVs are split into different numbers of portfolios, each with the same number of EVs, ranging from $70$ portfolios (each containing $20$ EVs) to a single large portfolio that includes all $1400$ EVs. The rationale behind this analysis is to explore whether there is a synergy effect by aggregating EVs into one large portfolio, rather than grouping them into multiple smaller portfolios. Recall that we report ex-post, out-of-sample profits \eqref{eq:obj}, which are calculated using the $3$-fold cross-validation process described in Section \ref{cross}.

Overall, Fig. \ref{fig:Synergy_bundle} reveals a significant synergy effect, with larger portfolios consistently leading to higher earnings. This synergy effect is immediately noticeable when the portfolio size is increased from $20$ to $50$ EVs (i.e., reducing the number of portfolios from  $70$ to $28$). The underlying reason for this synergy is that, by consolidating EVs into larger portfolios, the variability and uncertainty in total EV consumption within the portfolio decrease. This reduction in uncertainty allows the aggregator to more effectively leverage the $\rm{P90}$ requirement, enabling the placement of larger reserve quantity bids per EV.

Although the mean profit continues to rise as portfolios grow larger, the rate of profit increase gradually declines. This suggests that, beyond a certain point, the benefits of further aggregation become less pronounced.

%
%

As expected,  Fig. \ref{fig:Synergy_bundle} shows that the ALSO-X algorithm yields a higher profit than the CVaR approximation, as the latter tends to provide a more conservative solution. Nevertheless, the profit obtained by ALSO-X is still significantly lower than that achieved by the oracle model with perfect foresight. This suggests that there is considerable potential to improve the performance of the proposed model by better capturing future consumption uncertainty using historical data. As mentioned earlier, one possible improvement is to optimally select a subset of historical samples that are most relevant to the target hour, thus incorporating a form of conditionality into the model. This would allow for a more accurate prediction of flexibility. Additionally, caution must be exercised when selecting which historical samples to include, especially if the underlying stochastic environment is non-stationary. Over-relying on outdated data could reduce the model’s accuracy, and adjustments may be necessary to account for changes in the system over time.

\begin{figure}[t!]
    \centering
    \includegraphics[width=1\columnwidth]{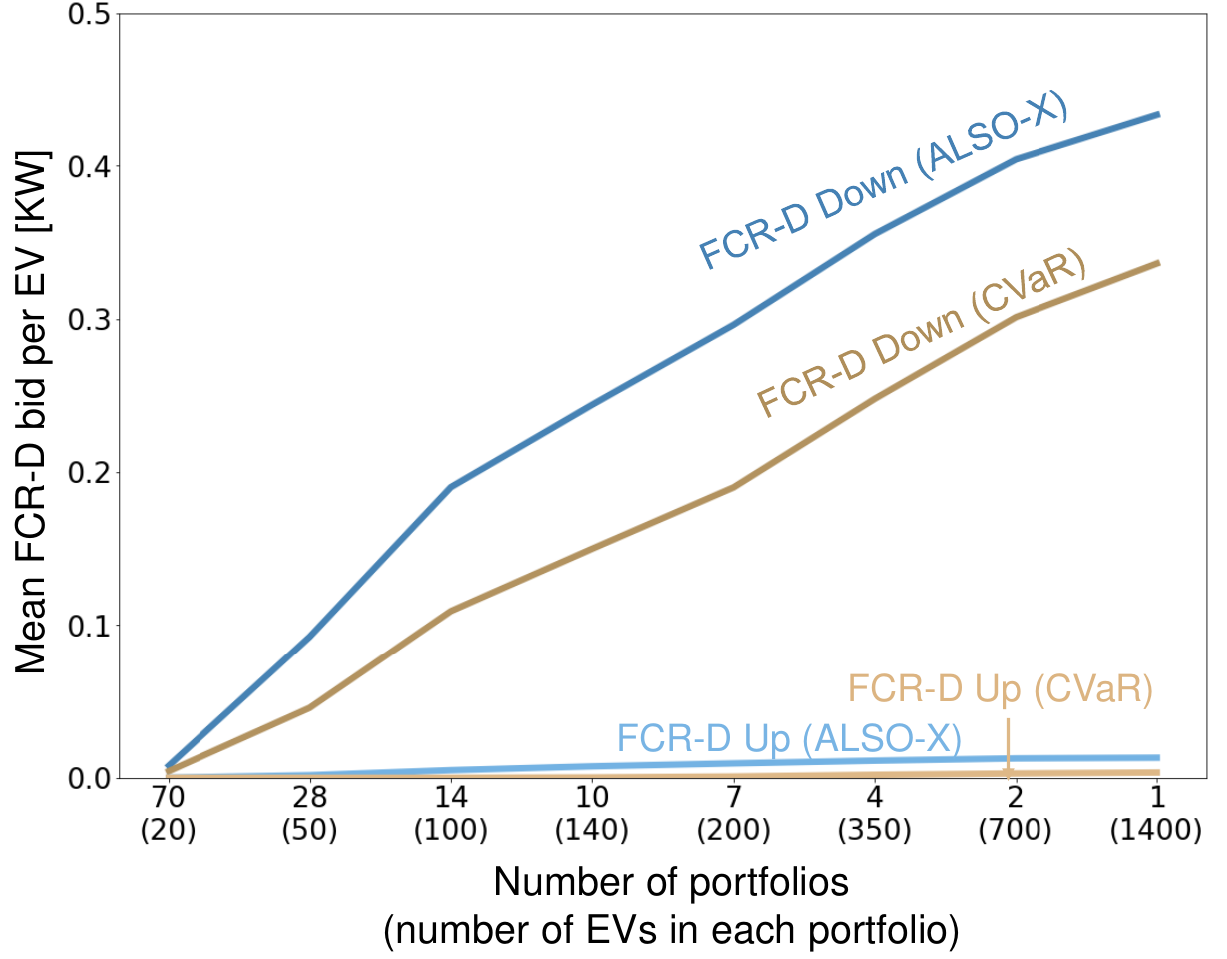}
    \caption{\small{The mean FCR-D bid size (kW) per EV.}}
\label{fig:bid_size}
\vspace{5mm}
\end{figure}

Similarly, Fig. \ref{fig:bid_size} shows the mean FCR-D bid size per EV. It is evident that the FCR-D down bid size (calculated using both the ALSO-X and CVaR methods) increases with portfolio size, demonstrating a clear synergy effect. A similar trend is observed for the size of FCR-D up bids, although these bids are consistently smaller than FCR-D down bids due to the more restrictive LER constraints.

\begin{table}[t]
    \centering
    \small
    \begin{tabular}{@{}cccc@{}}
        \toprule
                              & CVaR   & ALSO-X & Oracle \\ \midrule
        ${\rm{UBC}}^{\uparrow}$ (\%)   & 2.0  & 4.7  & 29.0 \\
        ${\rm{UBC}}^{\downarrow}$ (\%) & 10.8 & 15.7 & 49.9 \\ \bottomrule
    \end{tabular}
    \caption{\small{Average percentage of utilized bidding capacity, as defined in \eqref{sey1},  across $1400$ EVs for all portfolio sizes and hours.}}
    \label{tab:bid per}
    \vspace{2mm}
\end{table}

We now focus on the utilized bidding capacity as another metric of synergy, defined in \eqref{sey1}. Consistent with the results in Fig. 5, Table I shows that the oracle model bids a significantly higher proportion of the available flexibility. ALSO-X manages to sell more flexibility than CVaR, and thus outperforms it from a monetary perspective. However, all three models are still far from bidding $100$\% of their total flexibility.

In particular, it is noteworthy that the oracle model bids less than half of the total available flexibility. This is primarily due to the buffer imposed by the $\rm{LER}$ requirement. The LER constraint limits the amount of flexibility that can be bid, which reduces the total available flexibility for market bidding. From a system operator perspective, it would be interesting to explore how reducing the FCR-D market granularity to $30$ or even $15$ minutes could help optimize flexibility usage. Additionally, further optimization of the $\rm{LER}$ requirement could unlock more flexibility potential from stochastic assets, enabling a more efficient and dynamic participation in ancillary service markets. This could be particularly valuable in maximizing the reserve capacity provided by EV fleets and improving the overall system flexibility.

\vspace{1mm}
\subsection{Frequency of overbid}
First, we define a metric to assess compliance with the $\rm{P90}$, which measures how often bids exceed the realized flexibility. This metric, calculated ex-post, is referred to as the frequency of overbid (in \%), and is defined as:
\begin{align}\label{eq:freq-overbid}
    \text{Frequency of overbid} = \frac{1}{H\times N}\sum_{ m,h}y _{m,h},
\end{align}
where $y_{m,h}$ takes a value of one whenever an overbid occurs, and zero otherwise. In addition, $H$ denotes the number of hours and $N$ is the number of minutes being tested for the out-of-sample analysis. Without loss of generality, we will use this metric in a daily horizon, i.e., we will calculate how often the reserve shortfall occurs in every day of the historical $363$-day time period.

\begin{figure}[t]
    \centering
    \begin{subfigure}{.45\textwidth}
        \centering
        \includegraphics[width=\textwidth]{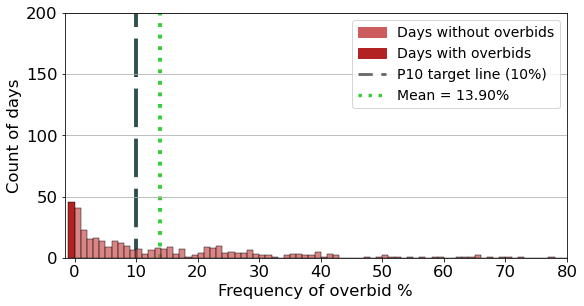}
        \caption{Naive approach}
        \label{fig:overbid_Naive}
    \end{subfigure}%
    \hfill
    \begin{subfigure}{.45\textwidth}
        \centering
        \includegraphics[width=\textwidth]{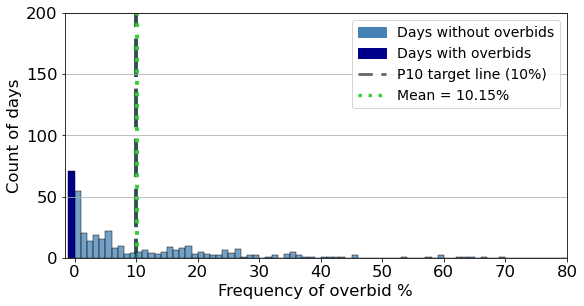}
        \caption{ALSO-X}
        \label{fig:overbid_ALSO_X}
    \end{subfigure}%
    \hfill
    \begin{subfigure}{.45\textwidth}
        \centering
        \includegraphics[width=\textwidth]{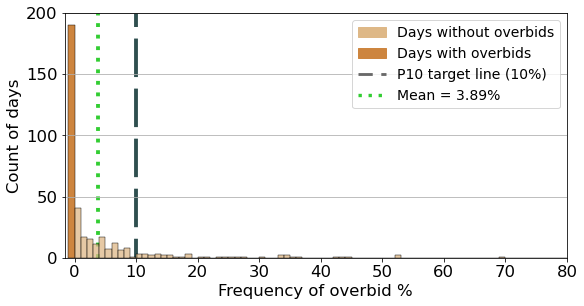}
        \caption{CVaR}
        \label{fig:overbid_CVAR}
    \end{subfigure}%
    \caption{\small{Frequency of overbids (i.e., reserve shortfall) tested for a portfolio of $1400$ EVs for $363$ days.}}
    \label{fig:general results 500}
        \vspace{-2mm}
\end{figure}

We now compare the naive approach, ALSO-X, and CVaR with respect to the compliance with the $\rm{P90}$ requirement. Fig. \ref{fig:general results 500} shows the daily frequency of overbid, as defined in \eqref{eq:freq-overbid}, for the portfolio of $1400$ EVs across all $363$ historical days. Recall that we use $242$ days for in-sample analysis and the remaining $121$ days for the ex-post out-of-sample analysis in a $3$-fold cross-validation process.

According to the middle plot of Fig. \ref{fig:general results 500}, the ALSO-X algorithm achieves the desired reserve shortfall rate of $10$\% almost perfectly, with the frequency of overbid across all simulated days at $10.15$\%. This must be regarded as a non-significant violation. In contrast the 10\% allowance rate is significantly exceeded by the naive approach, which operates with an overbid frequency of $13.9$\% (upper plot). On the other hand, the CVaR approximation (lower plot) shows a rate well below the threshold, with reserve overbidding occurring in only $3.89$\% of minutes. The CVaR results also show $190$ days with no overbidding, and a narrow spread among the days, whereas the spread for ALSO-X is broader.

\begin{figure}[t!]
    \centering
    \includegraphics[width=1\columnwidth]{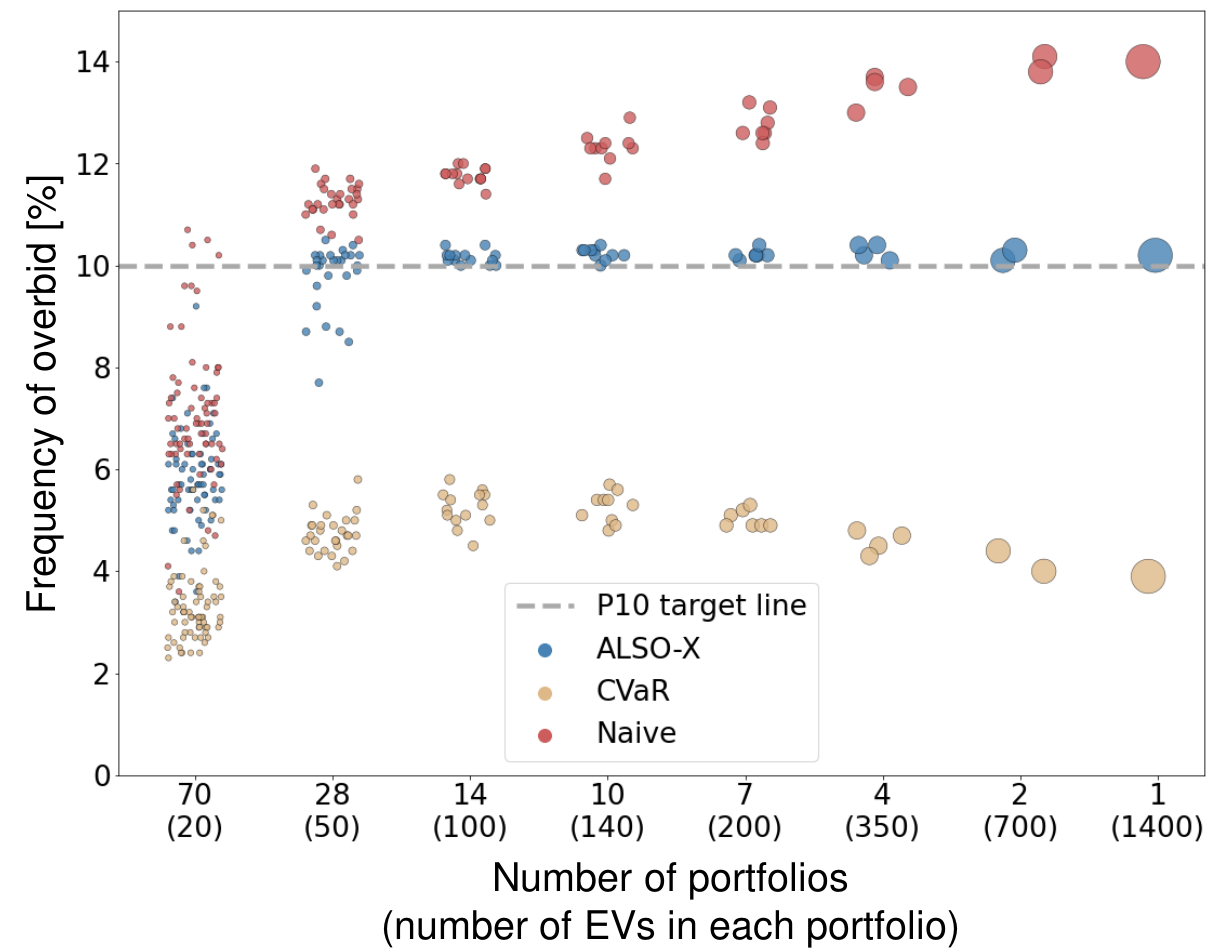}
    \caption{\small{The frequency of overbids across the tested $363$ days for different portfolios is shown. Each dot represents a unique portfolio containing a distinct set of the $1400$ EVs. The first cases on the x-axis correspond to portfolios where the $1400$ EVs are divided into $70$ portfolios, each containing $20$ EVs. The last case corresponds to the portfolio consisting of all $1400$ EVs, which is the same portfolio presented in Fig. \ref{fig:general results 500}.}}
\label{fig:bundle_overbids}
\end{figure}

Individual days with large overbidding are not problematic in itself according to the $\rm{P90}$ requirement, as it refers to the average violation rate over the entire period, which typically spans three months for Energinet. However, if the examination period were shorter than $363$ days, the naive approach and, to a lesser extent, the ALSO-X algorithm would be even more prone to violating the $\rm{P90}$ requirement, as there is a subset of days where the overbid frequency exceeds the $10$\% significantly. In contrast, CVaR offers more reliability in this regard. Moreover, while the experiment shown in Fig. \ref{fig:general results 500} uses a single portfolio, as an aggregator, one needs a bidding approach that is generalized and feasible across various portfolio compositions, as new EVs are continuously added. Fig. \ref{fig:bundle_overbids} shows the violation rates for each unique portfolio tested, in relation to profits. Generally, there is a lower violation rate for smaller portfolios, while CVaR operates well below $10$\% for all portfolio sizes. ALSO-X demonstrates the ability to meet the $10$\% target for nearly all portfolios with more than $50$ EVs, with only insignificant deviations above $10$\%. In contrast, the naive approach sees having an unstable tendency, steadily growing with larger portfolios, with most portfolios significantly violating the $10$\% allowance rate.

Overall, in the experiment conducted here, an aggregator using the naive approach would have been subject to exclusion from the FCR-D market. The approach show unstable and unreliable control of the frequency of overbid, making it an unviable solution for an EV aggregator. Thus, only CVaR and ALSO-X would be recommendable. Here, a trade-off between potential profit and adequacy exists. While the ALSO-X method may seem more attractive from a profit perspective, the CVaR method may be preferable in situations where adherence to the $\rm{P90}$ requirement is critical, as it is inherently conservative and provides a buffer against potential violations.

From a system operator's perspective, conservative approaches like CVaR reduce market liquidity, as they limit the reserve capacity that can be bid, but they also reduce uncertainty regarding supply. On the other hand, methods like ALSO-X, which fully exploit the rule set, may offer greater market liquidity but come with a higher risk of overbidding, leading to greater uncertainty in supply.

The key takeaway is that an accurate representation of the uncertainty in available flexibility ($F^{\uparrow}_{m,h}$, $F^{\downarrow}_{m,h}$, $F^{\rm{E}}_{m,h}$) is crucial when using methods like ALSO-X. In contrast, CVaR is more likely to produce feasible out-of-sample results that comply with the $\rm{P90}$ requirement, even when the uncertainty representation is less accurate. Therefore, the choice between these methods depends on the aggregator's priorities, balancing the need to maximize profit with ensuring compliance with grid requirements.

\section{Conclusion and future work}\label{sec:conclusion}

The frequency stability of the power grid is increasingly challenged by the growing penetration of renewable energy sources with their stochastic supply patterns. At the same time, demand flexibility, particularly from grid-connected assets such as EVs, holds significant untapped potential for grid balancing. In this study, we investigate how a portfolio of grid-connected EVs can be utilized to provide FCR-D services in Denmark, thus contributing to grid stability while allowing the aggregator to profit. Specifically, we analyze how the aggregator can operate within the framework of the newly implemented $\rm{P90}$ requirement set by the Danish system operator while ensuring compliance with the $\rm{LER}$ constraint.

We develop a joint chance-constrained optimization model to guide the bidding decision-making of the aggregator. This model is solved through two solution strategies: the ALSO-X algorithm, which iteratively solves a  linear program with relaxed binary variables, and the CVaR approximation, which offers a more conservative and linearized alternative.

Our analysis revealed a strong synergy effect from aggregating EVs. Larger portfolios of EVs enabled a higher quantity of flexibility for FCR-D bidding, resulting in increased profits for the aggregator and a more stable availability of flexibility. Notably, the CVaR method provided more conservative bidding results, bidding less aggressively and not fully exploiting the $10$\% reserve shortfall threshold, while the ALSO-X method leveraged the full flexibility potential. As a result, the latter method achieved higher profits but also carried a greater risk of overbidding.

A key limitation on the bidding capacity was the $\rm{LER}$ requirement, which requires aggregator to reserve $20$\% of their downwards bids in upwards flexibility. Future work could explore whether this constraint can be improved, as relaxing the $\rm{LER}$ requirement could allow the aggregator to bid more flexibility into the market, benefiting both the aggregator and the system operator. From the aggregator’s perspective, incorporating a more diverse set of technologies into the portfolio may help reduce the uncertainty in available flexibility, further improving the bidding decision-making. On the system operator's side, it would be valuable to analyze the trade-offs between supply security, market liquidity, and procurement costs when adjusting the $\rm{LER}$ requirement.



Another direction for future research is to assess the impact of the simplifications and assumptions used in this study. In particular, our approach to probabilistically modeling the available flexibility of the EV portfolio---based on historical data and simplified probabilistic representations---may be improved. The significant gap between the stochastic model and the oracle model, which assumes perfect foresight, suggests that more advanced forecasting methods could provide better results. We recommend that aggregators consider adopting advanced forecasting techniques to better capture the dynamics of flexibility availability. Additionally, distributionally robust optimization may offer a more reliable solution for cases where the empirical distribution of flexibility is misspecified.

\section*{Acknowledgement}

The authors would like to acknowledge Spirii for providing the data and Edoardo Simioni (Reel) for his valuable feedback. Our gratitude extends to Thomas Dalgas Fechtenburg (Energinet) for his guidance on current pre-qualifications and grid codes in Denmark. We also thank Torine Herstad Reed (DTU), Lesia Mitridati (DTU), and Bert Zwart (CWI Amsterdam) for insightful discussions. Finally, we would like to express our appreciation to the three anonymous reviewers for their constructive feedback.

\bibliographystyle{IEEEtran}
\bibliography{tex/bibliography/Bibliography}

\vfill

\end{document}